\documentclass[journal]{IEEEtran}
\UseRawInputEncoding
\usepackage{amsfonts,amssymb,amsmath,bm}
\usepackage{float}
\usepackage{graphicx}
\usepackage{subfigure}
\usepackage{cite,color}
\usepackage{cases}
\usepackage{epsfig}
\usepackage{verbatim}
\usepackage[ruled,linesnumbered]{algorithm2e}
\usepackage{dblfloatfix}
\usepackage{setspace}
\usepackage{multicol}
\usepackage{multirow}

\IEEEoverridecommandlockouts
\renewcommand\tabcolsep{2.0pt} 

\newtheorem{theorem}{\bf{Theorem}}
\newtheorem{lemma}{\textbf{Lemma}}

\begin{document}
\title{Channel Estimation for Hybrid Massive MIMO Systems with Adaptive-Resolution ADCs}
\author{Yalin Wang, Xihan Chen, Yunlong Cai, Benoit Champagne, and Lajos Hanzo
  \thanks{Y. Wang, X. Chen, and Y. Cai  are with the College of Information Science and Electronic Engineering, Zhejiang University, Hangzhou 310027, China (e-mail: wang\_yalin@zju.edu.cn; chenxihan@zju.edu.cn; ylcai@zju.edu.cn).
  B. Champagne is with the Department of Electrical and Computer Engineering, McGill University, Canada (e-mail: benoit.champagne@mcgill.ca).
  L. Hanzo is with the Department of ECS, University of Southampton, U.K. (e-mail: lh@ecs.soton.ac.uk).}
  }
	
\maketitle
\vspace{-3.3em}
\begin{abstract}
Achieving high channel estimation accuracy and reducing hardware cost
as well as power dissipation constitute substantial challenges in the
design of massive multiple-input multiple-output (MIMO) systems. To
resolve these difficulties, sophisticated pilot designs have been
conceived for the family of energy-efficient hybrid analog-digital
(HAD) beamforming architecture relying on adaptive-resolution
analog-to-digital converters (RADCs). In this paper, we jointly
optimize the pilot sequences, the number of RADC quantization bits and
the hybrid receiver combiner in the uplink of multiuser massive MIMO
systems. We solve the associated mean square error (MSE) minimization
problem of channel estimation in the context of correlated Rayleigh
fading channels subject to practical constraints. The associated
mixed-integer problem is quite challenging due to the nonconvex nature
of the objective function and of the constraints. By relying on
advanced fractional programming (FP) techniques, we first recast the
original problem into a more tractable yet equivalent form, which
allows the decoupling of the fractional objective function. We then
conceive a pair of novel algorithms for solving the resultant problems
for codebook-based and codebook-free pilot schemes, respectively. To reduce the design
complexity, we also propose a simplified algorithm for the codebook-based pilot scheme. Our
simulation results confirm the superiority of the proposed algorithms
over the relevant state-of-the-art benchmark schemes.
\end{abstract}
\IEEEpeerreviewmaketitle
\section{Introduction}
Massive multiple-input multiple-output (MIMO) systems rely on a large
number of base station (BS) antennas for simultaneously serving a few
dozens of users, while striking an attractive spectral efficiency (SE)
vs. energy efficiency (EE)
trade-off~\cite{massive_mimo1,massive_mimo2,massive_mimo3}. However, a
large number of antennas inevitably lead to an excessive
radio-frequency (RF) hardware cost and energy consumption. Hence, a
hardware-efficient hybrid analog-digital (HAD) beamforming
architecture has been proposed as an alternative to the fully digital
beamformer for practical implementation. Explicitly, this architecture
relies on an analog beamformer in the RF domain combined with a
low-dimensional digital beamformer in the baseband, hence allowing a
significant reduction in the number of RF chains required. This in
turn provides increased design flexibility for striking an attractive
performance vs. complexity trade-off.  The receiver's power dissipation
is dominated by that of the analog-to-digital converters (ADCs), since
it increases exponentially with the number of quantization
bits~\cite{exponential1},\cite{exponential2}. Hence, low-resolution
ADCs (LADCs) have been advocated for the RF chains of HAD receivers.

\vspace{-1.1em}
\subsection{Related Work}
Extensive research efforts have been invested in the design and performance evaluation of hybrid beamformers \cite{chongwen1,Letaief,yuwei,gao,zhuguangxu,partial-adaptive,fully-adaptive,dynamic}. Basically, the existing hybrid beamforming architectures can be mainly divided into the static partially-connected structure \cite{Letaief,yuwei,gao}, the static fully-connected structure \cite{Letaief,yuwei,zhuguangxu}, and the dynamic partially-connected or fully-connected structure \cite{partial-adaptive,fully-adaptive,dynamic}.
The authors of~\cite{Letaief} considered a single-user millimeter wave (mmWave) MIMO system and treated the hybrid beamforming weight design as a matrix factorization problem. Efficient alternating optimization algorithms were developed for both static partially-connected and static fully-connected structures.
As a further exploration, the authors of~\cite{yuwei} proposed heuristic algorithms for the design of static partially-connected and fully-connected HAD beamformers, permitting to maximize the overall SE of a broadband orthogonal frequency-division multiplexing (OFDM)-based system. Besides, the proposed algorithm in \cite{yuwei} for the fully-connected structure can achieve SE close to that of the optimal fully-digital solution with much less number of RF chains.
Moreover, to dynamically adapt to the spatial channel covariance matrix and improve the system performance, \cite{partial-adaptive,fully-adaptive,dynamic} proposed to design the dynamic-connected HAD beamforming architecture to realize a flexible analog beamforming matrix.
In \cite{partial-adaptive}, a dynamic sub-array approach was considered for OFDM systems, and a greedy algorithm to optimize the array partition based on the long-term channel characteristics was suggested.
Different from the partially-adaptive-connected structure in \cite{partial-adaptive}, the authors in \cite{fully-adaptive} proposed to implement the hybrid precoder with a fully-adaptive-connected structure. A joint optimization of switch-controlled connections and the hybrid precoders was formulated as a large-scale mixed-integer nonconvex problem with high dimensional power constraints. By modifying the on-off states of switch-controlled connections, this fully-adaptive-connected structure can realize a fully-connected structure or any possible sub-connected structure.
Furthermore, the HAD beamforming strategy has also been investigated in the context of novel relay-aided systems~\cite{relay_beamform,xuying} and in Terahertz communications~\cite{terahertz}.
Since employing LADCs in the massive MIMO regime has become indispensable for reducing the
power consumption and hardware cost, it has catalyzed substantial
interest in the recent literature. The authors of~\cite{ladc1}
analyzed the performance for transmission over flat-fading MIMO
channel using single-bit ADC and derived the capacity upper-bound both
at infinite and finite signal-to-noise ratios (SNR). The impact of the
spatial correlation of antennas on the rate loss caused by the coarse
quantization of LADCs was further studied in~\cite{ladc2}, where the
authors concluded that LADCs can achieve a sum rate performance much
closer to the case of ideal ADCs under spatially correlated channels.

On account of the benefits provided by HAD beamformers and LADCs, a number of studies have been proposed to characterize the performance of massive MIMO systems relying on the HAD beamforming architecture using LADCs.  The
pioneering contribution of~\cite{mojianhua} proposed a generalized
hybrid architecture using LADCs and verified that the achievable rate
is comparable to that obtained by high-precision ADC based receivers
at low and medium SNRs, which provides valuable insights for future
research. Intensive research efforts have also been dedicated to
analog/digital beamforming design~\cite{aspect_design1,aspect_design2},
to SE/EE optimization~\cite{aspect_opt1,aspect_opt2,aspect_opt3}, to
channel estimation~\cite{aspect_estimate} and to signal
detection~\cite{data_detector}.

However, previous research on HAD beamforming using LADCs has mainly
considered uniform quantizers having a fixed, predetermined number of
bits, which limited the performance of these systems due to coarse
quantization. As a further advance, it was shown that a
variable-resolution ADC or adaptive-resolution ADC (RADC) architecture
is preferable~\cite{radc_park,radc_choi}. In \cite{radc_park}, the
authors investigated a mixed-ADC structure designed for cloud radio
access networks (C-RAN). In particular, they developed an
ADC-resolution selection algorithm for maximizing either the SE or EE
based on an approximation of the generalized mutual information in the
low-SNR regime. In~\cite{radc_choi}, the authors developed a pair of
ADC bit allocation strategies for minimizing the quantization error
effects under a total ADC power constraint, thereby achieving an
improved performance. However, the separate design of the ADC
quantization bit allocation and hybrid beamforming matrices tends to
suffer from performance degradation. Hence, the authors
of~\cite{radc_icassp} jointly optimized both the on/off modes of the
RF processing chains and the number of ADC quantization bits.

As a further development, the authors of~\cite{radc_point} aimed for
jointly optimizing the sampling resolution of ADCs and the hybrid
beamforming matrices, which results in energy efficient solutions for
point-to-point mmWave MIMO systems. In light of~\cite{radc_point}, the
authors of~\cite{radc_shl} extended the joint design to multiuser
systems, hence achieving a significantly improved EE compared to the
existing schemes. The potential advantages of RADCs in the context of
various practical systems have also been reported
in~\cite{radc_schedule,radc_relay,radc_RIS}. The authors
of~\cite{radc_schedule} focused their attention on the uplink of
mmWave systems using RADCs and investigated the associated joint
resource allocation and user scheduling problem. In~\cite{radc_RIS}
the design of the reconfigurable intelligent surface (RIS) aided
mmWave uplink system relying on RADCs was investigated, demonstrating that an
RIS is capable of mitigating the performance erosion imposed by RADCs.

Nevertheless, the aforementioned studies are mainly based on perfect
instantaneous channel state information (CSI), which is assumed to be
known at the BS. In practice, the acquisition of perfect CSI cannot be
achieved in massive MIMO systems due to the inevitable channel
estimation errors~\cite{victor}. Therefore, how to efficiently design
the pilot signals for improving the precision of channel estimation is
of paramount importance. In practice a codebook-based pilot scheme is
preferred, where orthogonal pilot sequences are chosen from a given
codebook as a benefit of its low-complexity implementation and low
feedback overhead \cite{youli,kaiming_pilot}. However, allocating mutually orthogonal pilot
sequences to a large number of users for avoiding interference during
channel estimation would require excessive pilot lengths and their
orthogonality would still be destroyed upon convolution with the
dispersive channel impulse response (CIR). For this reason, the
carefully constructed reuse of a limited set of orthogonal pilot
sequences for different users for example is of paramount importance
for high-precision channel estimation. Further design alternatives
were proposed for massive MIMO systems for example
in~\cite{kaiming_pilot,pilot_ladc}, which dispense with a codebook,
hence they may be termed as codebook-free solutions.  However, they
tend to require a higher feedback overhead for attaining a high
channel estimation accuracy.
\begin{figure*}[t]
    \centering
    \includegraphics[width=0.7\textwidth]{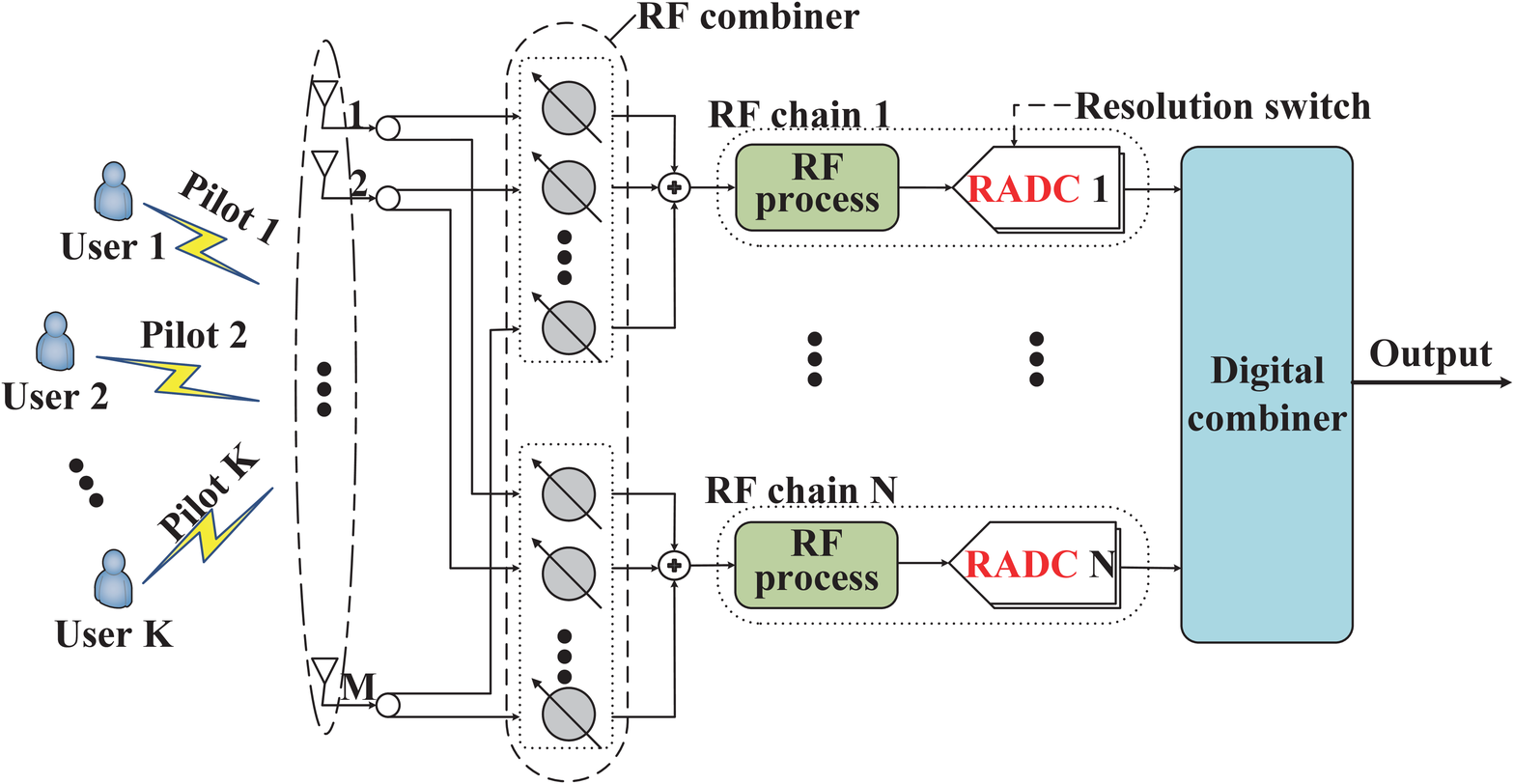}\\
    \caption{A multiuser massive MIMO uplink system adopting hybrid combiners with adaptive-resolution ADCs.}
    \label{fig_system_model}
\end{figure*}

\vspace{-1.1em}
\subsection{Main Contributions}
Despite the above advances, there is a paucity of research
contributions on jointly optimizing the pilot sequences, the number of
ADC quantization bits, and the HAD combiner for achieving
high-precision channel estimation in the uplink of a multiuser massive
MIMO system employing RADCs. Hence our inspiration is to fill this
knowledge-gap.  In particular, both codebook-based and codebook-free
pilot schemes are investigated, where for each scheme, we aim to minimize
the mean square error (MSE) of the channel estimate subject to a
transmit power constraint, to the constant-modulus constraint imposed
on the elements of the analog combining matrix, and to the additional
constraints on the number of quantization bits. In a nutshell, the
main contributions of this paper over the existing literature lie in
the following:
\begin{itemize}
\item[1)] We focus on jointly optimizing the pilot sequences, the HAD
  combiners and the RADC bit allocation in the presence of a
  correlated Rayleigh fading channel model. The channel estimation
  mean square error (MSE) minimization problem is formulated, which
  only requires the knowledge of channel statistics under practical
  operating conditions.

\item[2)] We first transform the highly nonconvex optimization problem
  into an equivalent but more tractable form by introducing auxiliary
  variables and employing fractional programming (FP)
  techniques. Then, we develop a new block coordinate descent
  (BCD) based algorithm for a codebook-free channel estimation
  scheme and a penalty dual decomposition (PDD) based algorithm for
  a codebook-based channel estimation scheme. Both of these
  iterative algorithms ensure convergence to the set of stationary
  solutions of the original optimization problem.  Furthermore, the
  computational complexity of the proposed algorithms is analyzed.
\item[3)] A simplified low-complexity algorithm is also presented for the codebook-based channel estimation scheme, which solves the MSE minimization problem  suboptimally but efficiently.

\item[4)] To characterize the benefits of our proposed algorithms, we
  provide exhaustive simulation results in terms of the MSE, sum rate
  and feedback overhead for a range of pertinent system settings.  We
  demonstrate that through the coordinated allocation of bits to the
  RADCs, the proposed algorithms can beneficially exploit the
  knowledge of channel statistics to accomplish the pilot
  design, while minimizing interference and improving the channel
  estimation accuracy.
\end{itemize}

\vspace{-1.1em}
\subsection{Organization and Notation}
This paper is organized as follows. In Section II and III, we
introduce the investigated system model and formulate the optimization
problem for the constrained channel estimation, respectively. In
Section IV and V, we propose efficient algorithms by solving the
formulated problems for the codebook-free and codebook-based pilot
schemes, respectively. Section VI provides simulation results to
appraise the performance of the proposed algorithms. The paper is
concluded in Section VII, whilst proofs and detailed derivations
appear in the Appendices.

\emph{Notations}: For a matrix $\mathbf{M}$, $(\mathbf{M})^{T}$,
$(\mathbf{M})^{*}$, $(\mathbf{M})^{H}$, and $\mathrm{vec}(\mathbf{M})$
denote its transpose, conjugate, conjugate transpose and
vectorization, respectively. $\mathbf{M}(i, j)$ denotes the element at
the intersection of row $i$ and column $j$. For a square matrix
$\mathbf{M}$, $\mathrm{Tr}(\mathbf{M})$, $\mathbf{M}^{-1}$, and $\|\mathbf{M}\|_{F}$
represents its trace, inverse and Frobenius norm, respectively.
$\mathrm{diag}(\mathbf{M})$ denotes a diagonal matrix consisting of
the diagonal elements of $\mathbf{M}$. $\mathbf{I}$ denotes an
identity matrix. For a vector $\mathbf{m}$,
$\mathrm{diag}(\mathbf{m})$ denotes a diagonal matrix with
$\mathbf{m}$ along its main diagonal and $\|\mathbf{m}\|$ denotes the
Euclidean norm of vector $\mathbf{m}$. The symbol $\otimes$ denotes
the Kronecker product. $\Re\{\cdot\}$ and $|\cdot|$ respectively
denote the real and magnitude parts of a complex number. $\lfloor
x\rfloor$ denotes the largest integer less than or equal to $x$ and
$\lceil x\rceil$ denotes the smallest integer greater than or equal to
$x$. We let $\mathbb{C}^{m\times n}$ ($\mathbb{R}^{m\times n}$) denote
$m \times n$ complex (real) space. $\mathbb{E}[\cdot]$ denotes the
expectation and $\mathcal{CN} (0, \sigma^{2})$ denotes the circularly
symmetric complex Gaussian distribution with mean 0 and variance
$\sigma^{2}$.

\section{System Model}
As shown in Fig. 1, we consider a multiuser massive MIMO uplink system
that adopts a static fully-connected hybrid AD combining structure with RADCs
at the BS. The BS which is equipped with $M > 1$ antennas and $N \ll
M$ RF chains, serves $K$ single-antenna users simultaneously. The
baseband output of each RF chain is fed to a dedicated RADC that
employs variable bit resolution to quantize the real and imaginary
parts of each analog signal. Moreover, we assume that the BS and users
are fully time-synchronized.

\vspace{-1.1em}
\subsection{Channel Model}
Without loss of generality, we consider a narrowband correlated
channel model. Let $\mathbf{h}_{k}\in \mathbb{C}^{M\times1}$ represent
the uplink channel from user $k\in\mathcal{K}\triangleq \{1,\ldots,
K\}$ to the BS. Then, the channel vector $\mathbf{h}_{k}$ can be
expressed as
\begin{equation}
\mathbf{h}_{k} = \mathbf{R}_{k}^{\frac{1}{2}}\mathbf{g}_{k},
\end{equation}
where $\mathbf{g}_{k}\in\mathbb{C}^{M\times 1}$ is a vector with independent and identically distributed (i.i.d.) elements distributed as $\mathcal{CN}(0,1)$, and $\mathbf{R}_{k} = \mathbb{E}[\mathbf{h}_{k}\mathbf{h}_{k}^{H}]$ denotes the channel covariance matrix for user $k$. $\mathbf{R}_{k}$ describes the spatial correlation properties of the channel
due to macroscopic effects of propagation, including path-loss and shadowing.
When the users are quasi-stationary, the path-loss and shadowing can be readily obtained based on the distance between the BS and user $k$ and stored at the BS as \emph{a priori} \cite{chenxihan},\cite{priori}.
\setcounter{equation}{8}
\begin{figure*}
\begin{align}\label{vec_h}
\mathrm{vec}(\mathbf{y}_{k}) = &(\mathbf{s}_{k}\otimes \mathbf{v}_{k}\mathbf{Q}_{\alpha}\mathbf{U})\mathbf{h}_{k} + \sum_{i\neq k}^{K}(\mathbf{s}_{i}\otimes \mathbf{v}_{k}\mathbf{Q}_{\alpha}\mathbf{U})\mathbf{h}_{i}\!
+\!(\mathbf{I}_{\tau}\otimes \mathbf{v}_{k}\mathbf{Q}_{\alpha}\mathbf{U})\mathrm{vec}(\mathbf{Z}) \!+\! (\mathbf{I}_{\tau}\otimes \mathbf{v}_{k})\mathrm{vec}(\mathbf{Z}_{q}).
\end{align}
\hrulefill
\end{figure*}
\setcounter{equation}{1}
\vspace{-1.1em}
\subsection{Pilot Sequences}
In this work, we focus on two types of pilot sequences, i.e., the
codebook-free and codebook-based pilots. For ease of exposition, we
denote as $\mathbf{s}_{k}\in \mathbb{C}^{\tau\times1}$ as the pilot
sequence transmitted by user $k\in\mathcal{K}$, where $\tau<K$ is the
length of the pilot sequence during each coherence interval. It is
noteworthy that $\tau$ is predetermined based on the coherence budget.

1) \emph{Codebook-free pilots}: As in previous works
\cite{kaiming_pilot},\cite{Yonina_Eldar_pilot}, we assume that each
pilot sequence $\mathbf{s}_{k}$ can be arbitrarily selected from the
$\tau$-dimensional space under the power constraint:
\begin{equation}\label{free_pilot}
\mathbf{s}_{k}\in \mathbb{C}^{\tau\times1} \quad\text{with}\quad \|\mathbf{s}_{k}\|^{2}\leq P_{k}^{\textrm{max}},
\end{equation}
where $P_{k}^{\textrm{max}}$ denotes the transmit power budget for user $k$.

2) \emph{Codebook-based pilots}: We denote the available codebook as
$\Upsilon= \{\boldsymbol{\upsilon}_{1}, \boldsymbol{\upsilon}_{2},
\cdots, \boldsymbol{\upsilon}_{\tau} \}$, where
$\boldsymbol{\upsilon}_{\iota}\in \mathbb{C}^{\tau\times 1}$ denotes
the $\iota$-th ($\iota\in\mathcal{T}\triangleq\{1,\ldots,\tau\}$)
potential pilot sequence. It is assumed that the different pilot
sequences meet the orthonormality conditions, i.e.,
$\boldsymbol{\upsilon}_{\iota}^{H}\boldsymbol{\upsilon}_{\iota'}=0$,
$\forall \iota\neq\iota'$ and
$\|\boldsymbol{\upsilon}_{\iota}\|^{2}=1,\forall \iota$. Then, the
pilot sequence of user $k$ is constructed as
\begin{equation}
\mathbf{s}_{k}=\sqrt{p_{k}}\boldsymbol{\varrho}_{k} \quad\text{with}\quad \|\mathbf{s}_{k}\|^{2}\leq P_{k}^{\textrm{max}},
\end{equation}
where $p_{k}$ denotes the transmit power of user $k$ and
$\boldsymbol{\varrho}_{k} \in \Upsilon$ denotes the codebook sequence
allocated to user $k$.


\vspace{-1.1em}
\subsection{Uplink Training}
In the uplink training phase, the BS estimates the uplink channels based on the pilot sequences simultaneously transmitted from the users.
Focusing on the $k$-th user, the received pilot signal at the BS can be expressed as
\begin{equation}
\mathbf{Y} = \mathbf{h}_{k}\mathbf{s}_{k}^{T} + \sum_{i\neq k}^{K}\mathbf{h}_{i}\mathbf{s}_{i}^{T} + \mathbf{Z},
\end{equation}
where the first term represents the desired contribution from user
$k$, the second term represents multiuser interference, and
$\mathbf{Z}\in\mathbb{C}^{M\times\tau}$ denotes the additive complex
Gaussian noise matrix with i.i.d. entries following the distribution
$\mathcal{CN}(0,\sigma^{2})$.

An analog combining matrix $\mathbf{U}\in\mathbb{C}^{N\times M}$ is
employed to process the received signal $\mathbf{Y}$ at the BS with
the goal of suppressing interference from the other users. In the HAD
architecture, the analog combiner is typically implemented using phase
shifters \cite{yuwei}, which imposes constant-modulus constraints on
the elements of the matrix $\mathbf{U}$. The output of the analog
combiner is given by
\begin{equation}
\bar{\mathbf{Y}} = \mathbf{U}(\mathbf{h}_{k}\mathbf{s}_{k}^{T} + \sum_{i\neq k}^{K}\mathbf{h}_{i}\mathbf{s}_{i}^{T} + \mathbf{Z}).
\end{equation}

We employ RADCs to quantize $\bar{\mathbf{Y}}$ as shown in
Fig. \ref{fig_system_model}, which enables flexible quantization bit
allocation for each baseband channel according to the radio
propagation characteristics. Such a refined design can efficiently
mitigate the quantization errors and greatly improve the system
performance with reduced hardware cost and power consumption. Let
integer $b_{n}$ denote the number of available quantization bits of
RADC $n$.  Assuming that the gain of automatic gain control is appropriately set, the additive quantization
noise model (AQNM) can be employed to reformulate the quantized signal \cite{radc_choi}. Then, based on the AQNM, the quantized output is specialized to
\begin{equation}
\mathbf{Y}_{q} = \mathcal{F}(\bar{\mathbf{Y}}) = \mathbf{Q}_{\alpha}\bar{\mathbf{Y}} + \mathbf{Z}_{q}, 
\end{equation}
where $\mathcal{F}(\cdot)$ is the element-wise quantization function,
$\mathbf{Q}_{\alpha}=\mathrm{diag}(\alpha_{1},\cdots,\alpha_{N})\in\mathbb{R}^{N\times
  N}$ is a diagonal gain matrix. Here, the quantization
gain $\alpha_{n}$ is a function of the number of quantization bit $b_{n}$
and defined as $\alpha_{n}=1
- \beta_{n}$, where $\beta_{n}$ is a normalized quantization error.
For $b_{n}\leq5$, $\beta_{n}$ can be expressed exactly in terms of
$b_n$ \cite{radc_choi}, while for $b_{n}>5$, they can be approximated
by $\beta_{n} \approx \frac{\pi\sqrt3}{2}2^{-2b_{n}}$. $\mathbf{Z}_{q}$
is the additive quantization noise which is independent of
$\bar{\mathbf{Y}}$. To facilitate
analytical derivations, we vectorize $\mathbf{Y}_{q}$ and obtain
$\mathrm{vec}(\mathbf{Y}_{q}) = (\mathbf{I}_{\tau}\otimes
\mathbf{Q}_{\alpha})\mathrm{vec}(\bar{\mathbf{Y}}) +
\mathrm{vec}(\mathbf{Z}_{q})$. $\mathrm{vec}(\mathbf{Z}_{q})$ obeys the complex Gaussian distribution with zero mean and  covariance matrix
\begin{align}
\mathbf{R}_{qq} &= \mathbb{E}[\mathrm{vec}(\mathbf{Z}_{q})\mathrm{vec}(\mathbf{Z}_{q})^{H}]\\\nonumber
&=(\mathbf{I}_{\tau}\otimes\mathbf{Q}_{\alpha})(\mathbf{I}_{\tau}\otimes\mathbf{Q}_{\beta})\mathrm{diag}(\mathbb{E}[\mathrm{vec}(\bar{\mathbf{Y}})\mathrm{vec}(\bar{\mathbf{Y}})^{H}]), \end{align}
where $\mathbf{Q}_{\beta}=\mathrm{diag}(\beta_{1},\cdots,\beta_{N})\in\mathbb{R}^{N\times N}$ and $\mathbb{E}[\mathrm{vec}(\bar{\mathbf{Y}})\mathrm{vec}(\bar{\mathbf{Y}})^{H}] = \sum_{i}(\mathbf{s}_{i}\otimes \mathbf{U})\mathbf{R}_{i}(\mathbf{s}_{i}\otimes \mathbf{U})^{H}+\sigma^{2}\mathbf{I}_{\tau}\otimes \mathbf{U}\mathbf{U}^{\mathbf{H}}$.

Finally, we leverage the digital processing techniques for quantization loss mitigation and interference cancellation.
Specifically, the retrieved signal of user $k$ at the output of the digital combiner $\mathbf{v}_{k}\in\mathbb{C}^{1\times N}$ is expressed as
\begin{equation}\label{signal_BS}
\mathbf{y}_{k} = \mathbf{v}_{k}\mathbf{Q}_{\alpha}\mathbf{U}(\mathbf{h}_{k}\mathbf{s}_{k}^{T} + \sum_{i\neq k}^{K}\mathbf{h}_{i}\mathbf{s}_{i}^{T} + \mathbf{Z}) + \mathbf{v}_{k}\mathbf{Z}_{q}.
\end{equation}

\vspace{-3mm}
\section{Problem Statement}
In this section, we first introduce the minimum MSE (MMSE)-based estimator and then formulate the problem under investigation.
\vspace{-2mm}
\subsection{MMSE Channel Estimation}
The BS aims to estimate the channel $\mathbf{h}_{k}$ based on the received pilot signal $\mathbf{y}_{k}$. To facilitate analytical derivations and achieve the estimation of the desired channel $\mathbf{h}_{k}$ using the MMSE estimator \cite{Yonina_Eldar_pilot}, we vectorize $\mathbf{y}_{k}$ in (\ref{signal_BS}) and obtain $\mathrm{vec}(\mathbf{y}_{k})$ shown at the top of this page.

Defining $\hat{\mathbf{h}}_{k}$ as the MMSE estimate of the channel $\mathbf{h}_{k}$, we have
\setcounter{equation}{9}
\begin{equation} \label{channel estimation}
\hat{\mathbf{h}}_{k} = \mathbf{A}_{k}^{H}\mathbf{B}_{k}^{-1}\mathrm{vec}(\mathbf{y}_{k}),
\end{equation}
where we define $\mathbf{x}_{k} \triangleq \mathbf{v}_{k}\mathbf{Q}_{\alpha}\mathbf{U}$, $\mathbf{Q}\triangleq\mathbf{Q}_{\alpha}\mathbf{Q}_{\beta}$,
$\mathbf{A}_{k}\triangleq \mathbf{s}_{k}\mathbf{x}_{k}\mathbf{R}_{k}$ and
$\mathbf{B}_{k}\triangleq \sum_{i}\mathbf{s}_{i}\mathbf{x}_{k}\mathbf{R}_{i}\mathbf{x}_{k}^{H}\mathbf{s}_{i}^{H} + \sigma^{2}\mathbf{x}_{k}\mathbf{x}_{k}^{H}\mathbf{I}_{\tau} + (\mathbf{I}_{\tau}\otimes \mathbf{v}_{k}\mathbf{Q})\mathrm{diag}\Big(\sum_{i}(\mathbf{s}_{i}\otimes \mathbf{U})\mathbf{R}_{i}(\mathbf{s}_{i}\otimes \mathbf{U})^{H}+\sigma^{2}\mathbf{I}_{\tau}\otimes \mathbf{U}\mathbf{U}^{\mathbf{H}}\Big)(\mathbf{I}_{\tau}\otimes \mathbf{v}_{k}^{H}).$
We note that matrix $\mathbf{B}_{k}$ is positive definite and therefore invertible. The corresponding MSE of user $k$ is given by
\begin{equation}\label{mse}
\mathrm{MSE}_{k} \triangleq \mathbb{E}[\|\hat{\mathbf{h}}_{k}-\mathbf{h}_{k}\|^{2}] = \mathrm{tr}(\mathbf{R}_{k}-\mathbf{A}_{k}^{H}\mathbf{B}_{k}^{-1}\mathbf{A}_{k}).
\end{equation}
The detailed derivations of $\hat{\mathbf{h}}_{k}$ and $\mathrm{MSE}_{k}$ are shown in Appendix A.
Then, the total MSE for the estimation of all the user channels can be expressed as
\begin{equation}\label{MSE}
\mathrm{MSE} = \sum_{k=1}^{K}\mathrm{MSE}_{k}=\sum_{k=1}^{K}\mathrm{tr}(\mathbf{R}_{k}-\mathbf{A}_{k}^{H}\mathbf{B}_{k}^{-1}\mathbf{A}_{k}).
\end{equation}

\subsection{Problem Formulation}
We note that the effectiveness of hybrid beamforming mainly depends on
the accuracy of CSI. In this work, we concentrate on the joint design
of the pilot sequence, HAD combiner, and the allocation of ADC
quantization bits, aiming to minimize the total MSE (\ref{MSE}) for
the given system model. To simplify notations, we introduce
$\mathbf{S}\triangleq[\mathbf{s}_{1},\cdots,\mathbf{s}_{K}]\in
\mathbb{C}^{\tau\times K}$,
$\mathbf{V}\triangleq[\mathbf{v}_{1}^{T},\cdots,\mathbf{v}_{k}^{T}]^{T}\in
\mathbb{C}^{K\times N}$, and
$\mathbf{b}\triangleq[b_{1},\cdots,b_{N}]^{T}\in \mathbb{R}^{N\times
  1}$.  Since the matrices $\mathbf{R}_{k}$ in (\ref{MSE}) do not
depend on the optimization variables
$\mathbf{b},\mathbf{U},\mathbf{V}$, and $\mathbf{S}$, minimizing the
$\mathrm{MSE}$ is equivalent to maximizing
$\sum_{k=1}^{K}\mathrm{tr}(\mathbf{A}_{k}^{H}\mathbf{B}_{k}^{-1}\mathbf{A}_{k})$. Hence,
we consider the following optimization problem
\begin{subequations} \label{P1}
\begin{align}
\underset{\mathbf{b},\mathbf{U},\mathbf{V},\mathbf{S}}{\max}\quad&f(\mathbf{b},\mathbf{U},\mathbf{V},\mathbf{S})\triangleq\sum_{k=1}^{K}\mathrm{tr}(\mathbf{A}_{k}^{H}\mathbf{B}_{k}^{-1}\mathbf{A}_{k}) \label{P1A}\\
\mbox{s.t.}\quad
&\|\mathbf{s}_{k}\|^{2}\leq P_{k}^{\textrm{max}},\quad\forall{k}, \label{P1B}\\
&|\mathbf{U}(n,m)| = \frac{1}{\sqrt{M}},\quad\forall{n,m}, \label{P1C}\\
&\check{b}_{n}\leq b_{n}\leq \hat{b}_{n},\quad\forall{n}, \label{P1D}\\
&\sum_{n=1}^{N}b_{n}\leq N\bar{b}, \label{P1E}
\end{align}
\end{subequations}
where $\check{b}_{n}$ and $\hat{b}_{n}$ respectively denote the minimum and maximum number of quantization bits ($\check{b}_{n}\leq \hat{b}_{n}$), $\bar{b}$ denotes the average number of quantization bits and $N\bar{b}$ is provided as the total budget of quantization bits at the BS. Constraint (\ref{P1C}) is imposed to enforce constant-modulus on the elements of analog combining matrix $\mathbf{U}$. Constraint (\ref{P1D}) limits the range of the quantization bits for each RADC, while constraint (\ref{P1E}) gives a threshold on the total ADC quantization bits at the BS. 

It should be emphasized that problem (\ref{P1}) is extremely difficult to solve due to the constant-modulus constraints, the nonconvex mixed-integer feasible set, and the highly nonconvex objective function with matrix ratio term $\mathbf{A}_{k}^{H}\mathbf{B}_{k}^{-1}\mathbf{A}_{k}$. To be specific, in the matrix fractional structure of the objective function, continuous variables $\mathbf{S},\mathbf{U},\mathbf{V}$ and discrete variable $\mathbf{b}$ appear in both the denominator and the numerator, which makes the problem intractable.

\vspace{-2mm}
\section{Proposed BCD-Based Algorithm for the Codebook-Free Pilot Scheme}
In this section, we focus on the codebook-free channel
estimation where the pilot sequences $\mathbf{s}_{k}$ meet condition
(\ref{free_pilot}). We first convert problem (\ref{P1}) into an
equivalent and mathematically tractable one based on the FP
method. Then, an efficient BCD-based joint design algorithm is
proposed to solve the resulting problem, where a series of subproblems
can be tackled via alternating optimization.

\vspace{-2mm}
\subsection{Problem Transformation}
With the aid of the advanced matrix FP techniques \cite{FP}, we employ the ratio-decoupling approach to transform problem (\ref{P1}) into a more tractable yet equivalent form. To this end, we first introduce the auxiliary variable $\mathbf{\Gamma}_{k}$ for each ratio term $\mathbf{A}_{k}^{H}\mathbf{B}_{k}^{-1}\mathbf{A}_{k}$.  Then problem
(\ref{P1}) can be converted into the equivalent problem
{\setlength\abovedisplayskip{2pt}
\setlength\belowdisplayskip{2pt}
\begin{subequations} \label{P2}
\begin{align}
\underset{\mathbf{b},\mathbf{U},\mathbf{V},\mathbf{S},\{\mathbf{\Gamma}_{k}\}}{\max}&f_{0}(\mathbf{b},\mathbf{U},\mathbf{V},\mathbf{S},\{\mathbf{\Gamma}_{k}\})\nonumber\\[-2mm]
&\triangleq\sum_{k=1}^{K}\mathrm{tr}(2\Re\{\mathbf{A}_{k}^{H}\mathbf{\Gamma}_{k}\}\!-\!\mathbf{\Gamma}_{k}^{H}\mathbf{B}_{k}\mathbf{\Gamma}_{k})\label{P2A}\\
\mbox{s.t.}\quad
&(\ref{P1B})-(\ref{P1E}),\\
&\mathbf{\Gamma}_{k}\in \mathbb{C}^{\tau\times M} ,\quad\forall{k}. \label{P2B}
\end{align}
\end{subequations}}It is observed that the constraint regarding to $\mathbf{\Gamma}_{k}$ in problem (\ref{P2}) are separable with respect to the other variables, i.e., $\mathbf{S},\mathbf{U},\mathbf{V},\mathbf{b}$.\footnote{Actually, matrix $\mathbf{V}$ is not involved in the constraints.}
When these variables are fixed, each auxiliary variable $\mathbf{\Gamma}_{k}$ can be optimally determined as follows:
{\setlength\abovedisplayskip{2pt}
\setlength\belowdisplayskip{2pt}
\begin{equation}\label{gamma}
\mathbf{\Gamma}_{k}^{\star} = \mathbf{B}_{k}^{-1}\mathbf{A}_{k}.
\end{equation}}The detailed proofs of the equivalence between problem (\ref{P1}) and
problem (\ref{P2}), as well as of the optimal solution (\ref{gamma})
for $\mathbf{\Gamma}_{k}$ are deferred to Appendix B. Using the matrix
quadratic transformation, where the cost function in (\ref{P1A}) is
replaced by that in (\ref{P2A}), we effectively decouple the numerator
and denominator in each term
$\mathbf{A}_{k}^{H}\mathbf{B}_{k}^{-1}\mathbf{A}_{k}$ and avoid the
difficulties posed by the nonconvex fractional objective function.

In light of (\ref{gamma}), it remains to optimize the other variables,
i.e., $\mathbf{S},\mathbf{U},\mathbf{V},\mathbf{b}$ , of the converted
problem (\ref{P2}). In the next subsection, we will specify how to
obtain optimal solutions for these variables.
\vspace{-3mm}
\subsection{Proposed BCD-Based Algorithm}
Note that the constraints (\ref{P1B})-(\ref{P1E}) in problem (\ref{P2}) are uncoupled with respect to the variables $\mathbf{b},\mathbf{U},\mathbf{V},\mathbf{S}$, i.e., each one of the constraints involve only one of these variables at a time. Hence, to reach a solution, we can decompose (\ref{P2}) into several independent subproblems each involving a single variable and solve problem (\ref{P2}) by means of the BCD algorithm. 
The corresponding developments are elaborated in further details below.

1) \emph{Optimization of $\mathbf{b}$}: In order to cope with the
difficulties posed by the discrete integer variables $b_{n}$, we first
relax $b_{n}$ into a continuous value $\tilde{b}_{n}$, solve the
resulting problem for $\tilde{b}_{n}$, and finally round each optimal
continuous value $\tilde{b}_{n}^{\star}$ to the nearest integer
\cite{radc_shl}. To determine the best integer quantization
bits and efficiently control quantization error, we employ the following criterion for
$\forall n$:
\begin{equation}\label{rounding}
b_{n}^{\star} = \left\{
\begin{aligned}
&\lfloor\tilde{b}_{n}^{\star}\rfloor,\quad \text{if}\quad \tilde{b}_{n}^{\star}- \lfloor{\tilde{b}_{n}^{\star}}\rfloor \leq \epsilon\quad(0\leq\epsilon\leq1),\\
&\lceil\tilde{b}_{n}^{\star}\rceil,\quad \text{otherwise}, \\
\end{aligned}
\right.
\end{equation}
where $\epsilon\in [0,1]$ is properly chosen so that
$\sum_{n=1}^{N}b_{n}^{\star}\leq N\bar{b}$ is satisfied. Notice that
simple rounding to $\lfloor\tilde{b}_{n}^{\star}\rfloor$ with
$\epsilon=1$ can always satisfy constraint (\ref{P1E}) and reduce the
power consumption while increasing the MSE and quantization errors. On the other hand,
rounding to $\lceil\tilde{b}_{n}^{\star}\rceil$ with $\epsilon=0$ may
violate the constraint (\ref{P1E}), i.e.,
$\sum_{n=1}^{N}b_{n}^{\star}\geq N\bar{b}$. Hence, considering
(\ref{rounding}), we apply the bisection method to find the optimal
value of $\epsilon$ \cite{round} and consequently, determine the
optimal allocation of the quantization bits, which greatly achieves the quantization error control.

We denote as
$\tilde{\mathbf{b}}=[\tilde{b}_{1},\cdots,\tilde{b}_{N}]^{T}\in
\mathbb{R}^{N\times 1}$ the vector of continuous variables after
relaxation, to be used in (\ref{P2}) in place of $\mathbf{b}$. With
the remaining variables being fixed, the subproblem for
$\tilde{\mathbf{b}}$ is formulated as
\begin{subequations}\label{sub_b}
\begin{align}
\underset{\tilde{\mathbf{b}}}{\max}\quad&f_{1}(\tilde{\mathbf{b}},\mathbf{U},\mathbf{V},\mathbf{S},\{\mathbf{\Gamma}_{k}\})\\
\mbox{s.t.}\quad&\check{b}_{n}\leq \tilde{b}_{n}\leq \hat{b}_{n},\quad\forall{n}, \label{sub_b_1}\\
&\sum_{n=1}^{N}\tilde{b}_{n}\leq N\bar{b}, \label{sub_b_2}
\end{align}
\end{subequations}
where the objective function is expressed as
\begin{align}\label{obj}
f_{1}&(\tilde{\mathbf{b}},\mathbf{U},\mathbf{V},\mathbf{S},\{\mathbf{\Gamma}_{k}\})\nonumber\\
=&\sum_{k}\bigg(\mathrm{tr}(2\Re\{\mathbf{R}_{k}\mathbf{x}_{k}^{H}\mathbf{s}_{k}^{H}\mathbf{\Gamma}_{k}\})-\!\sum_{i}\mathrm{tr}(\mathbf{\Gamma}_{k}^{H}\mathbf{s}_{i}\mathbf{x}_{k}\mathbf{R}_{i}\mathbf{x}_{k}^{H}\mathbf{s}_{i}^{H}\mathbf{\Gamma}_{k})\! \nonumber\\ &\!-\!\sum_{i}\mathrm{tr}(\mathrm{diag}(\mathbf{s}_{k}\mathbf{s}_{k}^{H})\mathbf{\Gamma}_{k}\mathbf{\Gamma}_{k}^{H})\mathbf{v}_{k}\mathbf{Q}\mathrm{diag}(\mathbf{U}\mathbf{R}_{i}\mathbf{U}^{H})\mathbf{v}_{k}^{H}\nonumber\\
&-\!\sigma^{2}\mathrm{tr}(\mathbf{\Gamma}_{k}\mathbf{\Gamma}_{k}^{H})(\mathbf{v}_{k}\mathbf{Q}\mathrm{diag}(\mathbf{U}\mathbf{U}^{H})\mathbf{v}_{k}^{H}\!+\!\mathbf{x}_{k}\mathbf{x}_{k}^{H})\bigg).
\end{align}

It is rather challenging to globally solve a nonconcave optimization
problem such as (\ref{sub_b}), due to the nature of the objective
function. To address this difficulty, we therefore resort to
successively solving a sequence of strongly concave approximate
problems. Specifically, by virtue of the successive concave
approximation (SCA) method \cite{radc_schedule}, we construct a
concave surrogate function in lieu of the objective function
(\ref{obj}) at each iteration. The surrogate function used at
iteration $t$ takes the form
\begin{equation}\label{surrogate_sca}
\bar{f}^{t}(\tilde{\mathbf{b}})\!=\!f_{1}(\tilde{\mathbf{b}}^{t})\!+\!\nabla_{\tilde{\mathbf{b}}}^{T}f_{1}(\tilde{\mathbf{b}}^{t})(\tilde{\mathbf{b}}\!-\!\tilde{\mathbf{b}}^{t})\!-\!\zeta\|\tilde{\mathbf{b}}\!-\!\tilde{\mathbf{b}}^{t}\|^{2},
\end{equation}
where $\nabla_{\tilde{\mathbf{b}}}f_{1}(\tilde{\mathbf{b}}^{t})$ is the gradient of $f_{1}(\tilde{\mathbf{b}})$ with respect to $\tilde{\mathbf{b}}$ at the current point $\tilde{\mathbf{b}}^{t}$ (which is calculated based on the chain rule), $\zeta$ is a positive constant, and the term $\zeta\|\tilde{\mathbf{b}}-\tilde{\mathbf{b}}^{t}\|^{2}$ is used to ensure the strong
concavity of $\bar{f}^{t}(\tilde{\mathbf{b}})$.
Therefore, at the $t$-th iteration of the SCA algorithm, we need to solve the following linearly constrained quadratic surrogate problem to update $\tilde{\mathbf{b}}^{t+1}$:
\begin{subequations} \label{sub_b1}
\vspace{-0.5em}
\begin{align}
\tilde{\mathbf{b}}^{t+1}=&\arg\underset{\tilde{\mathbf{b}}}{\max}\quad\bar{f}^{t}(\tilde{\mathbf{b}}) \\
&\mbox{s.t.}\quad(\ref{sub_b_1}),(\ref{sub_b_2}),
\end{align}
\end{subequations}whose solution can be efficiently obtained using the off-the-shelf CVX solver \cite{cvx}. Accordingly, we can find the optimal $\tilde{\mathbf{b}}^{\star}$ in an iterative fashion, and finally apply the procedure in (\ref{rounding}) to obtain the desired integer solution $\mathbf{b}^{\star}$.

2) \emph{Optimization of $\mathbf{U}$}: By keeping the other variables
fixed, the subproblem for $\mathbf{U}$ becomes a quadratic
optimization problem with constant-modulus constraints, which is given
by
\begin{subequations}\label{sub_U}
\begin{align}
\underset{\mathbf{U}}{\max}&\quad f_{1}(\tilde{\mathbf{b}},\mathbf{U},\mathbf{V},\mathbf{S},\{\mathbf{\Gamma}_{k}\}) \\[-2mm]
\mbox{s.t.}&\quad |\mathbf{U}(n,m)| = \frac{1}{\sqrt{M}},\forall{n,m}.
\end{align}
\end{subequations}
To handle the nontrivial constant-modulus constraints resulting from
the analog combiner, we use the one-iteration BCD-type algorithm
\cite{BCD_type} to recursively solve problem (\ref{sub_U}). The
detailed derivation is shown in Appendix C.


3) \emph{Optimization of $\mathbf{V}$}:
Similarly, the corresponding unconstrained subproblem for $\mathbf{V}$ is given by
\begin{equation} \label{sub_delta}
\underset{\mathbf{V}}{\max}\quad f_{1}(\tilde{\mathbf{b}},\mathbf{U},\mathbf{V},\mathbf{S},\{\mathbf{\Gamma}_{k}\})=\sum_{k}f_{1_{k}}(\tilde{\mathbf{b}},\mathbf{U},\mathbf{V},\mathbf{S},\{\mathbf{\Gamma}_{k}\}),
\end{equation}
where each term in the sum only involve the corresponding vector $\mathbf{v}_{k}$. Therefore, problem (\ref{sub_delta}) can be further decomposed into a sequence of simple per-user cases, each one being a quadratic optimization problem. These per-user subproblems can be efficiently solved by the first order condition $\partial f_{1_{k}}/\partial v_{k} = 0$. Specifically, the optimal value of $\mathbf{v}_{k}$
can be derived as 
\begin{align}
\mathbf{v}_{k}^{\star}=&\mathbf{s}_{k}^{H}\mathbf{\Gamma}_{k}\mathbf{R}_{k}\mathbf{U}^{H}\mathbf{Q}_{\alpha}\Bigg(\sum_{i}\mathbf{s}_{i}^{H}\mathbf{\Gamma}_{k}\mathbf{\Gamma}_{k}^{H}\mathbf{s}_{i}\mathbf{R}_{i}\mathbf{U}^{H}\mathbf{Q}_{\alpha}\nonumber\\&+\sigma^{2}\mathrm{tr}(\mathbf{\Gamma}_{k}\mathbf{\Gamma}_{k}^{H})(\mathbf{Q}_{\alpha}\mathbf{U}\mathbf{U}^{H}\mathbf{Q}_{\alpha}\!+\!\mathbf{Q}\mathrm{diag}(\mathbf{U}\mathbf{U}^{H}))\nonumber\\
&+\sum_{i}\mathrm{tr}(\mathrm{diag}(\mathbf{s}_{k}\mathbf{s}_{k}^{H})\mathbf{\Gamma}_{k}\mathbf{\Gamma}_{k}^{H})\mathbf{Q}\mathrm{diag}(\mathbf{U}\mathbf{R}_{i}\mathbf{U}^{H})\Bigg)^{-1}.\label{solution_delta}
\end{align}

4) \emph{Optimization of $\mathbf{S}$}: We now turn to the
optimization of the pilot matrix $\mathbf{S}$ while fixing the other
variables
$\mathbf{U},\mathbf{V},\tilde{\mathbf{b}},\{\mathbf{\Gamma}_{k}\}$. In
this case, the key step is to rewrite the objective function
(\ref{P2A}) in a quadratic form with respect to $\mathbf{s}_{k}$,
which results into
\begin{align}
f_{2}&(\tilde{\mathbf{b}},\mathbf{U},\mathbf{V},\mathbf{S},\{\mathbf{\Gamma}_{k}\}) \nonumber\\
=&\sum_{k}\mathrm{tr}(2\Re\{\mathbf{\Gamma}_{k}\mathbf{R}_{k}\mathbf{x}_{k}^{H}\mathbf{s}_{k}^{H}\})-\sum_{k}\mathbf{s}_{k}^{H}\mathbf{T}_{k}\mathbf{s}_{k} \nonumber\\ &-\sum_{k}\sum_{i}\mathrm{tr}(\mathbf{\Gamma}_{i}^{H}\mathbf{s}_{k}\mathbf{x}_{i}\mathbf{R}_{k}\mathbf{x}_{i}^{H}\mathbf{s}_{k}^{H}\mathbf{\Gamma}_{i}) + \mathrm{c},
\end{align}
where
\begin{equation}
\mathbf{T}_{k} = \sum_{r=1}^{N}
\begin{pmatrix}
\mathbf{c}_{k}^{r}\mathbf{d}_{11}^{r} & \cdots & \mathbf{c}_{k}^{r}\mathbf{d}_{1\tau}^{r}\\
\vdots & \ddots & \vdots\\
\mathbf{c}_{k}^{r}\mathbf{d}_{\tau1}^{r} & \cdots & \mathbf{c}_{k}^{r}\mathbf{d}_{\tau\tau}^{r}\\
\end{pmatrix}
\end{equation}
with $\mathbf{c}_{k}^{r}$ being the $r$th row vector of the matrix $\mathbf{C}_{k}=\mathbf{U}\mathbf{R}_{k}\mathbf{U}^{H}$, and $\mathbf{d}_{lj}^{r}$ being the $l$th $N\times1$ vector on the $[r+(j-1)N]$ column of $\mathbf{D} = \sum_{i}\mathrm{diag}((\mathbf{I}_{\tau}\otimes \mathbf{v}_{i}^{H})\mathbf{\Gamma}_{i}\mathbf{\Gamma}_{i}^{H}(\mathbf{I}_{\tau}\otimes \mathbf{v}_{i}\mathbf{Q}))$, and $\mathrm{c}$ is a constant term independent of variable $\mathbf{S}$.
Therefore, the subproblem for the codebook-free pilot matrix $\mathbf{S}$ is given by
\begin{align}
\underset{\mathbf{S}}{\max}\quad&f_{2}(\tilde{\mathbf{b}},\mathbf{U},\mathbf{V},\mathbf{S},\{\mathbf{\Gamma}_{k}\}) \nonumber\\
\mbox{s.t.}\quad&(\ref{P1B}),\label{P3A}
\end{align}
which can be solved by the Lagrange multiplier method. By associating
the Lagrange multiplier $\lambda_{k}$ to the corresponding power
budget constraint $\|\mathbf{s}_{k}\|^{2}\leq P_{k}^{\textrm{max}}$,
the Lagrange function for (\ref{P3A}) is given by
\begin{equation}
\mathcal{L}(\mathbf{S},\boldsymbol\lambda) = f_{2}(\tilde{\mathbf{b}},\mathbf{U},\mathbf{V},\mathbf{S},\{\mathbf{\Gamma}_{k}\}) - \sum_{k}\lambda_{k}(\|\mathbf{s}_{k}\|^{2}-P_{k}^{\textrm{max}}).
\end{equation}
By examing the first order optimality condition for $\mathcal{L}(\mathbf{S},\boldsymbol\lambda)$, the optimal value of $\mathbf{s}_{k}$ can be derived as
\setlength{\floatsep}{0.2cm}
\begin{algorithm}[tp]
\caption{Proposed BCD-Based Algorithm (\emph{\textbf{RADC Codebook-Free Scheme}})}\label{algorithm_non}
\textbf{Initialization:} Set all the variables to feasible values and define the tolerance of accuracy $\varepsilon_{1} > 0$\;
\Repeat{the increment on the value of the objective function in (\ref{P1A}) is less than $\varepsilon_{1} > 0$}{
Update the auxiliary variable $\mathbf{\Gamma}_{k}$'s according to (\ref{gamma})\;
Update variable $\tilde{\mathbf{b}}$ by solving (\ref{sub_b})\;
Update variable $\mathbf{U}$ by solving (\ref{sub_U})\;
Update variable $\mathbf{v}_{k}$'s (i.e., $\mathbf{V}$) according to (\ref{solution_delta})\;
Update variable $\mathbf{s}_{k}$'s (i.e., $\mathbf{S}$) according to (\ref{solution_S}) along with the Lagrangian multiplier $\lambda_{k}$'s in (\ref{lagrange})\;
}
\textbf{Rounding:} Round each optimal continuous value $\tilde{b}_{n}^{\star}$ according to (\ref{rounding}) and obtain the optimal $\mathbf{b}^{\star}$.
\end{algorithm}
\begin{equation}\label{solution_S}
\mathbf{s}_{k}^{\star} = \left(\sum_{i}\mathbf{x}_{i}\mathbf{R}_{k}\mathbf{x}_{i}^{H}\mathbf{\Gamma}_{i}\mathbf{\Gamma}_{i}^{H}+\mathbf{T}_{k}+\lambda_{k}\mathbf{I}_{\tau}\right)^{-1}\mathbf{\Gamma}_{k}\mathbf{R}_{k}\mathbf{x}_{k}^{H},
\end{equation}
where $\lambda_{k}\geq0$ should be optimally determined as
\begin{equation}\label{lagrange}
\lambda_{k}^{\star} = \left\{
\begin{aligned}
&0,\quad \text{if}\ \|\mathbf{s}_{k}\|^{2}\leq P_{k}^{\textrm{max}},\\
&\lambda_{k}>0\ \text{with}\ \|\mathbf{s}_{k}\|^{2} = P_{k}^{\textrm{max}},\ \text{otherwise}, \\
\end{aligned}
\right.\forall {k}.
\end{equation}

Note that (\ref{solution_S}) and (\ref{lagrange}) can be readily
solved via the bisection search; ultimately, we substitute the optimal
$\lambda_{k}^{\star}$ into (\ref{solution_S}) and obtain the optimal
$\mathbf{s}_{k}^{\star}$.

The corresponding BCD-based algorithm is summarized in Algorithm \ref{algorithm_non}.

\subsection{Convergence Analysis and Computational Complexity}
This subsection establishes the local convergence of Algorithm
\ref{algorithm_non} to stationary solutions and presents its detailed
computational complexity analysis. First, we introduce a key lemma,
which can be readily proved according to \cite{FP}.
\begin{lemma}\label{lemma1}
The objective functions  $f(\cdot)$ (\ref{P1A}) and $f_{0}(\cdot)$ (\ref{P2A}) satisfy
{\setlength\abovedisplayskip{2pt}
\setlength\belowdisplayskip{2pt}
\begin{equation}
f(\mathbf{b},\mathbf{U},\mathbf{V},\mathbf{S}) \geq f_{0}(\mathbf{b},\mathbf{U},\mathbf{V},\mathbf{S},\{\mathbf{\Gamma}_{k}\}),
\end{equation}}
with equality if and only if $\mathbf{\Gamma}_{k}$ satisfies (\ref{gamma}).
\end{lemma}

By invoking Lemma \ref{lemma1}, the convergence of Algorithm
\ref{algorithm_non} can be demonstrated. This property is summarized
as Theorem \ref{theorem1} below, and its proof can be found in
Appendix D.
\begin{theorem}\label{theorem1}
Algorithm \ref{algorithm_non} is guaranteed to converge, with the objective function $f(\mathbf{b},\mathbf{U},\mathbf{V},\mathbf{S})$ monotonically nondecreasing after each iteration. The solution after convergence is a stationary point of problem (\ref{P1}).
\end{theorem}

In the following, we analyze the computational complexity of Algorithm
\ref{algorithm_non}. We use the number of multiplications as a measure
of complexity and assume that $M\gg N\geq K$. Updating the auxiliary
variables $\mathbf{\Gamma}_{k}$ in (\ref{gamma}) involves the
calculation of $\mathbf{A}_{k}$, $\mathbf{B}_{k}$ and the inverse of
$\mathbf{B}_{k}$ based on Gauss-Jordan elimination with an overall
complexity
$\mathcal{O}(M^{2}K^{2}N\tau+MK^{2}N^{2}\tau^{2})$. According to the
proposed one-iteration BCD type algorithm \cite{BCD_type}, updating
all the entries of $\mathbf{U}$ once has a complexity of
$\mathcal{O}(M^{2}N^{2})$. Furthermore, the overall computational
complexity of $\mathbf{V}$ is on the order of
$\mathcal{O}(M^{2}K^{2}N+MK^{2}N^{2})$. The computational complexity
of optimizing $\mathbf{b}$ is dominated by computing the Jacobian
matrix of
$\sum_{k}\nabla_{\tilde{\mathbf{b}}}f_{1_{k}}(\tilde{\mathbf{b}}^{t})$
with respect to $\tilde{\mathbf{b}}$. Thus, the complexity for
updating $\mathbf{b}$ is $\mathcal{O}(I_{b}(M^{2}KN+MKN^{2}))$, where
$I_{b}$ denotes the number of iterations of the SCA method. As for the
pilot matrix $\mathbf{S}$, using the bisection method to search each
Lagrangian parameter $\lambda_{k}$ requires
$\log_{2}(\frac{\vartheta_{0}}{\vartheta_{s}})$ iterations to achieve
a desired accuracy, where $\vartheta_{0}$ is the initial interval size
and $\vartheta_{s}$ is the tolerance. Hence, the overall computational
complexity of updating $\mathbf{S}$ over all the users is
$\mathcal{O}(K\tau\log_{2}(\frac{\vartheta_{0}}{\vartheta_{s}})+M^{2}NK+MK^{2}\tau^{2})$. By
retaining dominant terms, the overall complexity of the proposed
Algorithm \ref{algorithm_non} is
$\mathcal{O}(I_{1}(M^{2}(N^{2}+K^{2}N\tau)+MK^{2}N^{2}\tau^{2}+K\tau\log_{2}(\frac{\vartheta_{0}}{\vartheta_{s}})))$, where $I_{1}$ denotes the number of iterations.

\vspace{-3mm}
\section{Proposed PDD-Based Algorithm for the Codebook-Based Pilot Scheme}
In this section, we focus on the codebook-based channel estimation
where pilot sequences are chosen from the codebook $\Upsilon$. Under
this setup, we first recast the corresponding problem (\ref{P2}) from
Section IV into a resource allocation problem with discrete binary
codeword indicator variables. Subsequently, we introduce a set of
auxiliary variables and propose an innovative PDD-based algorithm to
solve the optimization problem.
\setcounter{equation}{35}
\begin{figure*}
\begin{align}\label{sub_eta_k}
\underset{\boldsymbol\eta_{\iota}}{\max}\quad&\sum_{k}\bigg(\sqrt{p_{k}}\mathbf{e}_{k}^{T}\boldsymbol\eta_{\iota}\mathrm{tr}(2\Re\{\mathbf{\Gamma}_{k}\mathbf{R}_{k}\mathbf{x}_{k}^{H}\boldsymbol{\upsilon}_{\iota}^{H}\}) \!-\!p_{k}\boldsymbol\eta_{\iota}^{T}\mathbf{e}_{k}\mathbf{e}_{k}^{T}\boldsymbol\eta_{\iota}\boldsymbol{\upsilon}_{\iota}^{H}\mathbf{T}_{k}\boldsymbol{\upsilon}_{\iota}-p_{k}\mathbf{e}_{k}^{T}\boldsymbol\eta_{\iota}(\boldsymbol{\upsilon}_{\iota}^{H}\mathbf{T}_{k}\mathbf{a}+\mathbf{a}^{H}\mathbf{T}_{k}\boldsymbol\upsilon_{\iota}) \nonumber\\
&\!-\!\sum_{i}p_{k}\boldsymbol\eta_{\iota}^{T}\mathbf{e}_{k}\mathbf{e}_{k}^{T}\boldsymbol\eta_{\iota}\mathrm{tr}(\mathbf{\Gamma}_{i}^{H}\boldsymbol{\upsilon}_{\iota}\mathbf{x}_{i}\mathbf{R}_{k}\mathbf{x}_{i}^{H}\boldsymbol{\upsilon}_{\iota}^{H}\mathbf{\Gamma}_{i}) -\sum_{i}p_{k}\mathbf{e}_{k}^{T}\boldsymbol\eta_{\iota}\mathrm{tr}(2\Re\{\mathbf{\Gamma}_{i}^{H}\boldsymbol{\upsilon}_{\iota}\mathbf{x}_{i}\mathbf{R}_{k}\mathbf{x}_{i}^{H}\mathbf{a}^{H}\mathbf{\Gamma}_{i}\})\nonumber\\
&-\frac{1}{2\rho}\big(|\boldsymbol{\eta}_{\iota}^{T}\mathbf{e}_{k}-\hat\eta_{k\iota}+\rho\tilde{\lambda}_{k\iota }|^{2}+|\boldsymbol{\eta}_{\iota}^{T}\mathbf{e}_{k}(\hat\eta_{k\iota}-1)+\rho\bar{\lambda}_{k\iota}|^{2}
+|\sum_{\iota}\boldsymbol{\eta}_{\iota}^{T}\mathbf{e}_{k}-1+\rho\hat{\lambda}_{k}|^{2}\big)\bigg),
\end{align}
\hrulefill
\end{figure*}
\begin{figure*}
\begin{align}\label{solution_eta}
\boldsymbol\eta_{\iota}^{\star}=&\bigg(\sum_{k}\big(2\rho\sum_{i}p_{k}\mathrm{tr}(\mathbf{\Gamma}_{i}^{H}\boldsymbol{\upsilon}_{\iota}\mathbf{x}_{i}\mathbf{R}_{k}\mathbf{x}_{i}^{H}\boldsymbol{\upsilon}_{\iota}^{H}\mathbf{\Gamma}_{i})+2\rho p_{k}\boldsymbol{\upsilon}_{\iota}^{H}\mathbf{T}_{k}\boldsymbol{\upsilon}_{\iota}+(\hat\eta_{k\iota}-1)^{2}+2\big)\mathbf{e}_{k}\mathbf{e}_{k}^{T}\bigg)^{-1}\nonumber\\
&\times\Big(\sum_{k}\big(\rho\sqrt{p_{k}}\mathrm{tr}(2\Re\{\mathbf{\Gamma}_{k}\mathbf{R}_{k}\mathbf{x}_{k}^{H}\boldsymbol{\upsilon}_{\iota}^{H}\})-\rho(\tilde{\lambda}_{k\iota}+\bar{\lambda}_{k\iota}\hat\eta_{k\iota}-\bar{\lambda}_{k\iota}+\hat\lambda_{k})-\sum_{\iota'\neq\iota}\boldsymbol\eta_{\iota'}^{T}\mathbf{e}_{k}\nonumber\\
&+\hat\eta_{k\iota}+1-\rho\sum_{i}p_{k}\mathrm{tr}(2\Re\{\mathbf{\Gamma}_{i}^{H}\boldsymbol{\upsilon}_{\iota}\mathbf{x}_{i}\mathbf{R}_{k}\mathbf{x}_{i}^{H}\mathbf{a}^{H}\mathbf{\Gamma}_{i}\})-\rho p_{k}(\boldsymbol{\upsilon}_{\iota}^{H}\mathbf{T}_{k}\mathbf{a}+\mathbf{a}^{H}\mathbf{T}_{k}\boldsymbol\upsilon_{\iota})\big)\mathbf{e}_{k}\Big).
\end{align}
\hrulefill
\end{figure*}
\setcounter{equation}{30}
\vspace{-0.5cm}
\subsection{Proposed PDD-Based Algorithm}
Since $\mathbf{s}_{k}$ is herein structured as
$\mathbf{s}_{k}=\sqrt{p_{k}}\boldsymbol{\varrho}_{k}$ where
$\boldsymbol{\varrho}_{k}$ is chosen from the codebook $\Upsilon$, the
codebook-based pilot design can be regarded as a pilot resource
allocation problem. Consequently, how to allocate the limited
orthogonal pilot in $\Upsilon$ to the different users to reduce the
channel estimation error is of great importance. We introduce $\eta_{k
  \iota}\in\{0,1\}$ as the allocation indicator, where
$\eta_{k\iota}=1$ signifies that the best orthogonal pilot
$\boldsymbol{\upsilon}_{\iota}$ is assigned to user $k$; otherwise, we
have $\eta_{k \iota}=0$. Hence, using $\eta_{k\iota}$ we can write
$\boldsymbol{\varrho}_{k}=\sum_{\iota=1}^{\tau}\eta_{k\iota}\boldsymbol{\upsilon}_{\iota}$
and
$\mathbf{s}_{k}=\sqrt{p_{k}}\sum_{\iota=1}^{\tau}\eta_{k\iota}\boldsymbol{\upsilon}_{\iota}$.

With each $\mathbf{s}_{k}$ expressed in terms of  $(p_{k},\eta_{k\iota})$, problem (\ref{P2}) can be equivalently converted to the following problem:
\begin{subequations} \label{P_allocation}
\begin{align}
\underset{\mathcal{Z},\{\eta_{k\iota}\}}{\max}&f_{3}(\mathcal{Z},\{\eta_{k\iota}\})\!=\!f_{0}(\mathbf{b},\mathbf{U},\mathbf{V},\mathbf{S},\{\mathbf{\Gamma}_{k}\})\mid_{\mathbf{s}_{k}=\sqrt{p_{k}}\sum_{\iota=1}^{\tau}\eta_{k\iota}\boldsymbol{\upsilon}_{\iota}}\label{P_allocation_A}\\
\mbox{s.t.}\quad
&(\ref{P1B})-(\ref{P1E}), (\ref{P2B}),\\
&\sum_{\iota=1}^{\tau}\eta_{k\iota}=1, \forall k,\label{P_allocation_B}\\
&\eta_{k\iota}\in \{0,1\}, \forall k,\iota, \label{P_allocation_C}
\end{align}
\end{subequations}
where $\mathcal{Z}\triangleq \{\mathbf{U},\mathbf{V},\mathbf{b},\mathbf{p},\{\mathbf{\Gamma}_{k}\}\}$  with $\mathbf{p}=[p_{1},\ldots,p_{K}]^{T}$ represents the search variables, and constraint (\ref{P_allocation_B}) guarantees that each user is associated with a single pilot sequence.

To address the difficulty posed by discrete binary constraints
(\ref{P_allocation_B}) and (\ref{P_allocation_C}), we introduce the
auxiliary variables $\{\hat\eta_{k\iota}\}$, in terms of which these
constraints can be equivalently expressed as
\begin{equation}\label{binary_constraint}
\quad0\leq\hat\eta_{k\iota}\leq1,
\end{equation}
\begin{equation}\label{equal_constraint}
\sum_{\iota=1}^{\tau}\boldsymbol{\eta}_{\iota}^{T}\mathbf{e}_{k}=1, \quad\boldsymbol{\eta}_{\iota}^{T}\mathbf{e}_{k}=\hat\eta_{k\iota}, \quad\boldsymbol{\eta}_{\iota}^{T}\mathbf{e}_{k}(\hat\eta_{k\iota}-1)=0,
\end{equation}
where $\boldsymbol{\eta}_{\iota}\triangleq[\eta_{1\iota},\cdots,\eta_{K\iota}]^{T}$ and $\mathbf{e}_{k}$ is the $k$-th
column of identity matric $\mathbf{I}_{K}$. Then problem (\ref{P_allocation}) is then equivalent to
\begin{subequations} \label{P_allocation1}
\begin{align}
\underset{\mathcal{Z},\{\eta_{k\iota}\},\{\hat\eta_{k\iota}\}}{\max}\quad&f_{3}(\mathcal{Z},\{\eta_{k\iota}\})\label{P_allocation1_A}\\
\mbox{s.t.}\quad
&(\ref{P1B})-(\ref{P1E}), (\ref{P2B}),(\ref{binary_constraint}),(\ref{equal_constraint}),
\end{align}
\end{subequations}
where importantly, the variables $\eta_{k\iota}$ are no longer limited
to binary values. To solve problem (\ref{P_allocation1}), we next
introduce the proposed PDD-based algorithm which exhibits a
double-loop structure to solve problem (\ref{P_allocation1}).  Based
on the PDD framework \cite{pdd,pdd1}, we first add a penalized version
of the equality constraints in (\ref{equal_constraint}) to the
objective function (\ref{P_allocation1_A}), thereby obtaining the
following augmented Lagrangian (AL) problem
\begin{subequations} \label{AL_problem}
\begin{align}
\underset{\mathcal{Z},\{\eta_{k\iota}\},\{\hat\eta_{k\iota}\}}{\max}&f_{3}(\mathcal{Z},\{\eta_{k\iota}\})-\frac{1}{2\rho}\sum_{k=1}^{K}\sum_{\iota=1}^{\tau}|\boldsymbol{\eta}_{\iota}^{T}\mathbf{e}_{k}-\hat\eta_{k\iota}+\rho\tilde{\lambda}_{k\iota }|^{2}\nonumber\\
&-\frac{1}{2\rho}\sum_{k=1}^{K}\sum_{\iota=1}^{\tau}|\boldsymbol{\eta}_{\iota}^{T}\mathbf{e}_{k}(\hat\eta_{k\iota}-1)+\rho\bar{\lambda}_{k\iota}|^{2}\nonumber\\
&-\frac{1}{2\rho}\sum_{k=1}^{K}|\sum_{\iota=1}^{\tau}\boldsymbol{\eta}_{\iota}^{T}\mathbf{e}_{k}-1+\rho\hat{\lambda}_{k}|^{2}\label{AL_problem_A}\\
\mbox{s.t.}\quad
&(\ref{P1B})-(\ref{P1E}), (\ref{P2B}),(\ref{binary_constraint}),
\end{align}
\end{subequations}
where
$\{\tilde{\lambda}_{k\iota}\},\{\bar{\lambda}_{k\iota}\},\{\hat{\lambda}_{k}\}$,
$k\in\mathcal{K}$, $\iota\in\mathcal{T}$ denote the Lagrange
multipliers and $\rho\in\mathbb{R}_{+}$ denotes the penalty
coefficient. We note that problems (\ref{P_allocation1}) and
(\ref{AL_problem}) are equivalent in the limit $\rho \rightarrow 0$,
which is at the hearth of the PDD method. Specifically, in the
PDD-based algorithm, the inner loop solves the AL problem with fixed
AL multipliers and penalty coefficient, while the outer loop aims to
update the dual variables while reducing the penalty coefficient in
light of the constraint violation.

Since the constraints in problem (\ref{AL_problem}) are separable, we
can address the AL problem (\ref{AL_problem}) in the inner loop with
the BCD method. Particularly, the subproblems for $\mathbf{b}$,
$\mathbf{U}$ and $\mathbf{V}$ are the same as problems (\ref{sub_b}),
(\ref{sub_U}) and (\ref{sub_delta}) discussed in Section IV-B,
respectively, and can therefore be solved using the same methods. The
optimization of the remaining variables in the inner loop,
i.e. $\{\boldsymbol\eta_{\iota}\}, \{\hat\eta_{k\iota}\}$, and
$\mathbf{p}$, is explained in further detail below.

1) \emph{Optimization of $\{\boldsymbol\eta_{\iota}\}$}: We optimize $\{\boldsymbol\eta_{\iota}\}$ in parallel for $\iota = 1,...,\tau$, with the remaining variables being fixed. The subproblem of optimizing $\boldsymbol\eta_{\iota}$ can be simplified as the unconstrained problem in (\ref{sub_eta_k}) shown at the top of this page, where $\mathbf{a}\triangleq\sum_{\iota'\neq\iota}\boldsymbol\eta_{\iota'}^{T}\mathbf{e}_{k}\boldsymbol\upsilon_{\iota'}$. By examining the first-order optimality condition of (\ref{sub_eta_k}), we derive the closed-form solution of $\boldsymbol\eta_{\iota}$ shown in (\ref{solution_eta}) at the top of this page. Finally, we use the one-iteration BCD method to update $\{\boldsymbol\eta_{\iota}\}$ based on (\ref{solution_eta}).

2) \emph{Optimization of $\{\hat\eta_{k\iota}\}$}: We optimize
$\{\hat\eta_{k\iota}\}$ in parallel with the other variables
fixed. The corresponding subproblem for $\hat\eta_{k\iota}$ can be
expressed as
\setcounter{equation}{37}
\begin{align}\label{sub_hat_eta}
\underset{\hat\eta_{k\iota}}{\min}\quad&\frac{1}{2\rho}(|\boldsymbol{\eta}_{\iota}^{T}\mathbf{e}_{k}-\hat\eta_{k\iota}+\rho\tilde{\lambda}_{k\iota }|^{2}+|\boldsymbol{\eta}_{\iota}^{T}\mathbf{e}_{k}(\hat\eta_{k\iota}-1)+\rho\bar{\lambda}_{k\iota}|^{2})\nonumber\\
\mbox{s.t.}\quad
&(\ref{binary_constraint}).
\end{align}
Problem (\ref{sub_hat_eta}) features a scalar quadratic objective
function of $\hat\eta_{k\iota}$, for which we can directly obtain the
unconstrained minimizer as
\begin{equation}\label{solution_hat_eta}
\hat\eta_{k\iota}^{'}=\frac{\rho\tilde{\lambda}_{k\iota}+\eta_{k\iota}-\rho\eta_{k\iota}\bar{\lambda}_{k\iota}+\eta_{k\iota}^{2}}{1+\eta_{k\iota}^{2}}.
\end{equation}
According to constraint (\ref{binary_constraint}), $\hat\eta_{k\iota}$ must satisfy $0\leq\hat\eta_{k\iota}\leq1$, we can obtain the optimal solution of the constrained problem (\ref{sub_hat_eta}) as follows:
\begin{equation}\label{opt_hat_eta}
\hat\eta_{k\iota}^{\star} = \left\{
\begin{aligned}
&0,\quad \hat\eta_{k\iota}^{'}\leq 0,\\
&\hat\eta_{k\iota}^{'},\quad 0<\hat\eta_{k\iota}^{'}<1, \\
&1,\quad\hat\eta_{k\iota}^{'}\geq 1,\\
\end{aligned}
\right.\quad\forall {k,\iota}.
\end{equation}

3) \emph{Optimization of $\mathbf{p}$}: By fixing the other variables, the corresponding subproblem for variable $p_{k}$ can be expressed as
\begin{align}\label{sub_p}
\underset{p_{k}}{\max}\quad f_{p_{k}}=&\sqrt{p_{k}}
\mathrm{tr}(2\Re\{\mathbf{\Gamma}_{k}\mathbf{R}_{k}\mathbf{x}_{k}^{H}\boldsymbol{\varrho}_{k}^{H}\}) \nonumber\\ &\!-\!p_{k}\sum_{i}\!\mathrm{tr}(\mathbf{\Gamma}_{i}^{H}\boldsymbol{\varrho}_{k}\mathbf{x}_{i}\mathbf{R}_{k}\mathbf{x}_{i}^{H}\boldsymbol{\varrho}_{k}^{H}\mathbf{\Gamma}_{i})\!-\!p_{k}\boldsymbol{\varrho}_{k}^{H}\mathbf{T}_{k}\boldsymbol{\varrho}_{k}\nonumber\\
\mbox{s.t.}\quad
&\|\sqrt{p_{k}}\|^{2}\leq P_{k}^{\textrm{max}},
\end{align}
where $\boldsymbol{\varrho}_{k}=\sum_{\iota=1}^{\tau}\eta_{k\iota}\boldsymbol{\upsilon}_{\iota}$. The variable $p_{k}$ can be determined uniquely by solving the first-order equation $\partial f_{p_{k}}/\partial{p_{k}}=0$, which yields
\begin{equation}\label{solution_p}
p_{k}^{\star}\!=\!\min\left\{P_{k}^{\textrm{max}},\left(\frac{\mathrm{tr}(\Re\{\mathbf{\Gamma}_{k}\mathbf{R}_{k}\mathbf{x}_{k}^{H}\boldsymbol{\varrho}_{k}^{H}\}) }{\!\sum_{i}\!\mathrm{tr}(\mathbf{\Gamma}_{i}^{H}\boldsymbol{\varrho}_{k}\mathbf{x}_{i}\mathbf{R}_{k}\mathbf{x}_{i}^{H}\boldsymbol{\varrho}_{k}^{H}\mathbf{\Gamma}_{i})\!+\!\boldsymbol{\varrho}_{k}^{H}\mathbf{T}_{k}\boldsymbol{\varrho}_{k}}\right)^{2}\right\}.
\end{equation}

In the outer iteration of the PDD-based algorithm, the dual variables
$\{\tilde{\lambda}_{k\iota}\},\{\bar{\lambda}_{k\iota}\},\{\hat{\lambda}_{k}\}$,
$\forall k\in\mathcal{K}$, $\iota\in\mathcal{T}$, can be updated
according to
\begin{subequations}\label{dual_variable}
\begin{align}
\tilde{\lambda}_{k\iota}^{r+1}=\tilde{\lambda}_{k\iota}^{r}+(\boldsymbol{\eta}_{\iota}^{T}\mathbf{e}_{k}-\hat\eta_{k\iota})/\rho^{r},\\
\bar{\lambda}_{k\iota}^{r+1}=\bar{\lambda}_{k\iota}^{r}+(\boldsymbol{\eta}_{\iota}^{T}\mathbf{e}_{k}(\hat\eta_{k\iota}-1))/\rho^{r},\\
\hat\lambda_{k}^{r+1}=\hat\lambda_{k}^{r}+(\sum_{\iota=1}^{\tau}\boldsymbol{\eta}_{\iota}^{T}\mathbf{e}_{k}-1)/\rho^{r},
\end{align}
\end{subequations}
where superscript $r$ refers to the iteration number of the outer
loop. As for the penalty parameter $\rho$, it is decremented according
to $\rho^{r+1}=a\rho^{r+1}$, where $0 < a < 1$.  To measure the
violation of the equality constraints, we adopt the constraint
violation indicator $\delta$ , which is defined as
\begin{equation}\label{hbar}
\delta=\underset{k,\iota}{\max}\left\{|\boldsymbol{\eta}_{\iota}^{T}\mathbf{e}_{k}-\hat\eta_{k\iota}|,|\boldsymbol{\eta}_{\iota}^{T}\mathbf{e}_{k}(\hat\eta_{k\iota}-1)|,|\sum_{\iota=1}^{\tau}\boldsymbol{\eta}_{\iota}^{T}\mathbf{e}_{k}-1|\right\}.
\end{equation}
When $\delta\leq\varepsilon_{2}$, where $\varepsilon_{2}$ denotes the
tolerance on the constraint violation, the algorithm terminates.  The
corresponding PDD-based algorithm is summarized in Algorithm
\ref{algorithm_or}.

\begin{algorithm}[htp]
\caption{Proposed PDD-Based Algorithm (\emph{\textbf{RADC Codebook-Based Scheme}})}\label{algorithm_or}
\textbf{Initialization:} Dual variables $\{\tilde{\lambda}_{k\iota},\bar{\lambda}_{k\iota},\hat{\lambda}_{k}\}^{0}$, primal variables $\{\mathbf{U},\mathbf{V},\mathbf{b},\mathbf{p},\{\eta_{k\iota}\},\{\hat\eta_{k\iota}\},\{\mathbf{\Gamma}_{k}\}\}^{0}$, tolerances $\varepsilon_{2} > 0$ and $\varepsilon_{3} > 0$, penalty factor $\rho^{0}>0$, $0<a<1$, $\hat\varepsilon^{0}>0$, $r = 0$ \;
\Repeat{the termination criterion is met: $\delta\leq \varepsilon_{2}$}{\Repeat{the increment on the value of the objective function in (\ref{AL_problem_A}) is less than $\varepsilon_{3}$}{
Update the auxiliary variable $\mathbf{\Gamma}_{k}$'s according to (\ref{gamma})\;
Update variable $\tilde{\mathbf{b}}$ by solving (\ref{sub_b})\;
Update variable $\mathbf{U}$ by solving (\ref{sub_U})\;
Update variable $\mathbf{v}_{k}$'s (i.e., $\mathbf{V}$) according to (\ref{solution_delta})\;
Update variable $\boldsymbol\eta_{\iota}$'s according to (\ref{solution_eta})\;
Update variable $\hat\eta_{k\iota}$'s according to (\ref{opt_hat_eta})\;
Update variable $\mathbf{p}$ according to (\ref{solution_p})\;}
Calculate the constraint violation $\delta$ according to (\ref{hbar})\;
\eIf{$\delta\leq \hat\varepsilon^{r}$}{Update dual variables according to (\ref{dual_variable})\;}{Set $\rho^{r+1}=a\rho^{r}$\;}
Set $\hat\varepsilon^{r+1} = a\delta$, and $r=r+1$\;
}
\textbf{Rounding:} Round each optimal continuous value $\tilde{b}_{n}^{\star}$ according to (\ref{rounding}) and obtain the optimal $\mathbf{b}^{\star}$.
\end{algorithm}

\vspace{-1.1em}

\subsection{Convergence Analysis and Computational Complexity}
Based on the discussion of \cite{pdd,pdd1}, the proposed PDD-based
Algorithm 2 is guaranteed to converge to a stationary point of problem
(\ref{P_allocation1}). Since there is no relaxation or approximation
during the transformation from the original problem in
(\ref{P_allocation}) to problem (\ref{P_allocation1}), problem
(\ref{P_allocation}) and problem (\ref{P_allocation1}) share the same
stationary solution. Therefore, Algorithm 2 is guaranteed to converge
to the stationary point of problem (\ref{P_allocation}).


As for the computational complexity of Algorithm \ref{algorithm_or}, the inner loop procedure is similar to Algorithm \ref{algorithm_non} except for the optimization of pilot sequences. The complexity of optimizing $\{\boldsymbol\eta_{\iota}\},\{\hat\eta_{k\iota}\}$ and $\mathbf{p}$ in Algorithm \ref{algorithm_or} is $\mathcal{O}((M^{2}N+MN)K\tau+K^{3})$. Thus, the overall computational
complexity of the proposed Algorithm \ref{algorithm_or} is
$\mathcal{O}(I_{2}I_{3}(M^{2}(N^{2}+K^{2}N\tau)+MK^{2}N^{2}\tau^{2}))$, where $I_{2}$ and $I_{3}$ are the numbers of iterations in the outer and inner loops, respectively.

\begin{algorithm}[ht]
\caption{Proposed Simplified Algorithm (\emph{\textbf{Simplified Scheme}})}\label{algorithm_simple}
\textbf{Initialization:} Set all the variables to feasible values and define the tolerance of accuracy $\varepsilon_{4} > 0$\;
\textbf{Apply} the SGPA method to achieve codebook-based pilot allocation \cite{youli}\;
\Repeat{the increment on the value of the objective function in (\ref{simpleA}) is less than $\varepsilon_{4} > 0$}{
Update the auxiliary variable $\mathbf{\Gamma}_{k}$'s according to (\ref{gamma})\;
Update variable $\tilde{\mathbf{b}}$ by solving (\ref{sub_b})\;
Update variable $\mathbf{U}$ by solving (\ref{sub_U})\;
Update variable $\mathbf{v}_{k}$'s (i.e., $\mathbf{V}$) according to (\ref{solution_delta})\;
Update variable $\mathbf{p}$ according to (\ref{solution_p})\;
}
\textbf{Rounding:} Round each optimal continuous value $\tilde{b}_{n}^{\star}$ according to (\ref{rounding}) and obtain the optimal $\mathbf{b}^{\star}$.
\end{algorithm}

\vspace{-0.5cm}
\section{Simplified Algorithm for the Codebook-Based Pilot Scheme}
\vspace{-0.1cm}
To reduce the computational complexity, we propose a simplified algorithm based on the statistical greedy pilot allocation (SGPA) method \cite{youli} for the codebook-based channel estimation. The main idea of the SGPA method is that the channel covariance matrices of the users who reuse the pilots should be as orthogonal as possible. We define the orthogonality between two channel covariance matrices as $\pi=\frac{\mathrm{tr}(\mathbf{R}_{k}^{H}\mathbf{R}_{k'})}{\|\mathbf{R}_{k}\|_{F}\|\mathbf{R}_{k'}\|_{F}} \in [0,1]$, where $\pi=0$ indicates that these two channel covariance matrices are orthogonal. Smaller $\pi$ means stronger orthogonality and weaker similarity. The SGPA process can be divided into two steps: 1) we first assign $\tau$ orthogonal pilots of the available codebook $\Upsilon$ to $\tau$ users with similar channel covariance matrices; 2) we then allocate the “best” pilot to each of the remained $K-\tau$ users so that the channel covariance matrices of the users who reuse the pilots are as orthogonal as possible. Additional details about SGPA method can be found in \cite[Algorithm 1]{youli}. Hence, we firstly employ the SGPA method to allocate pilot sequences among users to achieve pilot reuse. Then, with the allocated pilot $\boldsymbol{\varrho}_{k},\forall k$, the original problem  reduces to the following problem:
{\setlength\abovedisplayskip{0mm} \setlength\belowdisplayskip{0mm}
\begin{subequations} \label{simple}
\begin{align}
\underset{\mathcal{Z}}{\max}\quad&f_{4}(\mathcal{Z})=f_{0}(\mathbf{b},\mathbf{U},\mathbf{V},\mathbf{S},\{\mathbf{\Gamma}_{k}\})\mid_{\mathbf{s}_{k}=\sqrt{p_{k}}\boldsymbol{\varrho}_{k}}\label{simpleA}\\
\mbox{s.t.}\quad
&(\ref{P1B})-(\ref{P1E}), (\ref{P2B}).
\end{align}
\end{subequations}}We can solve this problem using the BCD approach with guaranteed convergence, and the updates of $\mathbf{U},\mathbf{V},\mathbf{b},\mathbf{p},\{\mathbf{\Gamma}_{k}\}$ are obtained using the same method shown in the PDD-based algorithm in Section V-A.

Overall, the simplified algorithm is presented in Algorithm \ref{algorithm_simple} with a complexity order of $\mathcal{O}(I_{4}(\!M^{2}N^{2}+M^{2}K^{2}N\tau+MK^{2}N^{2}\tau^{2})+M^{2}K^{3})$, where $I_{4}$ denotes the number of iterations. It is worth noting that the complexity of the simplified algorithm is much lower than that of the proposed PDD-based algorithm, since it only has one loop. However, its performance is not as good as the performance of the PDD-based algorithm, which will be verified by simulation results. Consequently, our proposed PDD-based algorithm for the codebook-based pilot scheme therefore offers a practical trade-off between complexity and performance, while the simplified algorithm for the codebook-based pilot scheme solves problem (\ref{P2}) suboptimally but with reduced complexity.

\section{Simulation Results}
In this section, simulations are conducted to validate the effectiveness of our proposed algorithms. We consider a single cell scenario with a cell radius $r_{d}=300$ m.
The BS, which is located in the center of the cell, is equipped with $M=64$ antennas and $N=12$ RF chains. A total of $K=12$ users are uniformly distributed within the cell area.
The pathloss of user $k$ is calculated as $30.6 + 36.7 \log_{10}(d_{k})$ in dB \cite{pathloss}, where $d_{k}$ is the distance in meters between that user and the BS. The log-normal shadow fading, i.e., the corresponding loss in dB, follows a Gaussian distribution with zero mean and variance $\sigma_{s}^{2}=8$ dB. We set the maximum transmit power level $P_{k}^{\mathrm{max}}$ as $20$ dBm and the system bandwidth as 10 MHz. The background noise power spectral density is set as $-169$ dBm/Hz \cite{noise}.
We adopt a geometry-based spatially correlated channel model with a half-wavelength space uniform linear array for simulations \cite{aspect_opt2}. Specifically, the channel vector between the BS and user $k$ is modeled as $\mathbf{h}_{k}=\sum_{i=1}^{L_{p}}\gamma_{k,i}\mathbf{a}(\theta_{k,i})$, where $L_{p} = 5$ is the number of channel paths for each user, $\mathbf{a}(\theta)$ is the array response vector with generic expression given by $\mathbf{a}(\theta)=\frac{1}{\sqrt{M}}[1,e^{j\pi\sin{\theta}},\ldots,e^{j\pi(N-1)\sin{\theta}}]$, $\theta_{k,i}$ are the angles of arrival, independently generated with a Laplace distribution with an angle spread $\sigma_{AS}=10$, and $\gamma_{k,i}$ are the complex path gains following the $\mathcal{CN}(0,\sigma_{k,i}^{2})$ distribution. The $\sigma_{k,i}^{2}$ are randomly generated from an exponential distribution and normalized such that $\sum_{i=1}^{L_{p}}\sigma_{k,i}^{2}=G_{k}$, where $G_{k}$ is the desired average channel gain. For simplicity, we set $\check{b}_{n}=\check{b}=1$, $\hat{b}_{n}=\hat{b}=8$, and $\bar{b}=3$ \cite{radc_shl}.

We utilize the normalized MSE (NMSE) to evaluate the channel estimation performance of the proposed schemes \cite{chongwen2}. Specifically, we define
$\mathrm{NMSE}_{k}=\frac{1}{N}\sum_{n=1}^{N}\frac{\|\hat{\mathbf{h}}_{k}^{n}-\mathbf{h}_{k}^{n}\|^2}{\|\mathbf{h}_{k}^{n}\|^{2}}$,
where $\hat{\mathbf{h}}_{k}^{n}$ is the MMSE estimate of the $k$-th user's channel $\mathbf{h}_{k}^{n}$ obtained in the $n$-th Monte Carlo trial and $N=1000$ is the total number of such trials, and we let 
$\mathrm{NMSE}=\sum_{k=1}^{K}\mathrm{NMSE}_{k}$.

For comparison, the following three benchmarks are also considered in the simulations:
\begin{itemize}
\item Random pilot (RP) scheme \cite{kaiming_pilot}: The pilot sequences are generated independently and randomly, with entries following the complex Gaussian distribution under the maximum power constraint.
\item Random allocation (RA) scheme \cite{kaiming_pilot}: A subset of $\tau$ users are randomly selected and assigned  mutually orthogonal pilots, while the remaining users are randomly allocated pilots in the codebook $\Upsilon$.
\item Uniform quantization (UQ) schemes \cite{aspect_opt2}: For both the codebook-free and codebook-based cases, LADCs with fixed and identical number of quantization bits are implemented at the BS (marked as `UQ codebook-free scheme' and `UQ codebook-based scheme'). These schemes serve as benchmarks to investigate how the RADCs influence the system performance.
\end{itemize}

\begin{figure}[t]
		\centering
		\includegraphics[width=8cm]{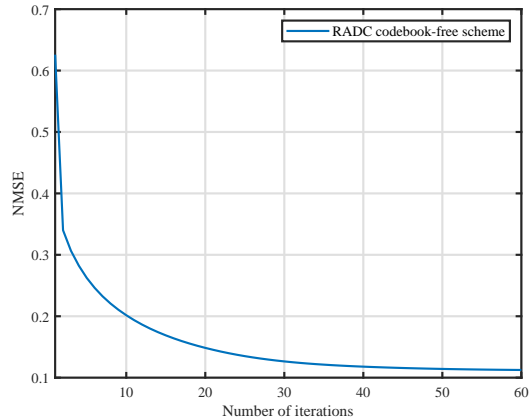}
		\caption{NMSE versus number of iterations for the proposed BCD-based algorithm ($M=64$, $K=12$, $\tau=8$, and $\bar{b}=3$).}
        \label{fig_converge1}
\end{figure}

\begin{figure}[t]
		\centering		\includegraphics[width=8cm]{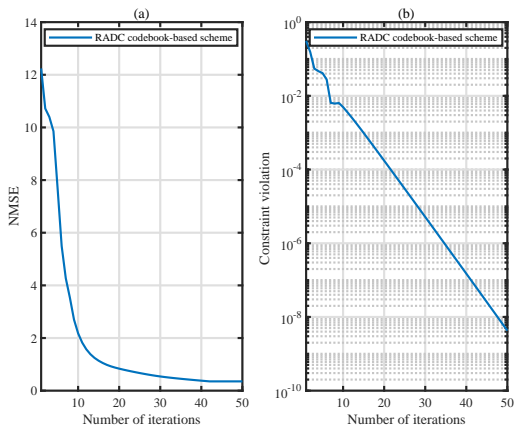}
		\caption{NMSE (a) and constraint violation (b) versus number of outer iterations for the proposed PDD-based algorithm ($M=64$, $K=12$, $\tau=8$, and $\bar{b}=3$).}
        \label{fig_converge2}
\end{figure}

Let us commence by examining the convergence behavior of the proposed
algorithms. The NMSE performance versus number of iterations for the proposed
BCD-based algorithm and PDD-based algorithm are presented in
Fig. \ref{fig_converge1} and Fig. \ref{fig_converge2}(a); in addition,
Fig. \ref{fig_converge2}(b) shows the constraint violation for the proposed
PDD-based algorithm. It is observed that for both proposed algorithms, the
NMSE converges fast within a few iterations. In particular, it
indicates that the penalty terms in the PDD-based algorithm decrease to a value
below $10^{-8}$ after 50 iterations in
Fig. \ref{fig_converge2}(b). These results verify the ability of the
proposed algorithms to effectively handle problem (\ref{P1}).
\begin{figure}[t]
		\centering
		\includegraphics[width=8cm]{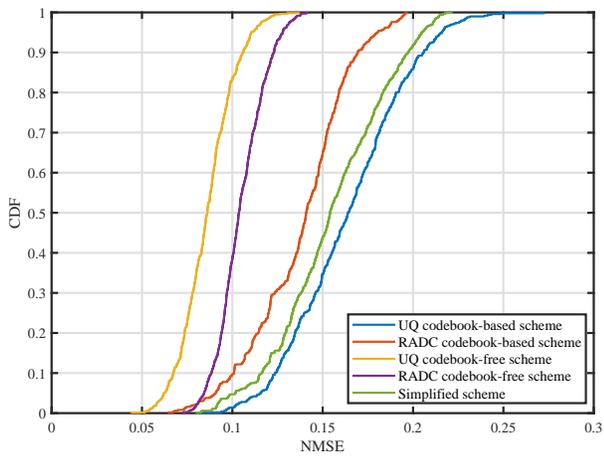}
		\caption{Cumulative distribution function of the NMSE ($M=64$, $K=12$, $\tau = 10$, and $\bar{b}=3$).}
        \label{cdf}
\end{figure}

In Fig. \ref{cdf}, we take a closer look at the cumulative
distribution function (CDF) of the NMSE for different
schemes. Remarkably, the proposed RADC codebook-free scheme
outperforms that of the other competing schemes at any percentile,
while the transition of its CDF from 0 to 1 occurs over a smaller
range. The results also illustrate that the RADC codebook-based scheme and the simplified scheme
yield better NMSE performance than the UQ codebook-based scheme. However, the simplified scheme achieves a suboptimal performance compared to the RADC codebook-based scheme, due to the fact that the pilots are heuristically allocated in advance in the simplified algorithm.

\begin{figure}[t]
		\centering
		\includegraphics[width=8cm]{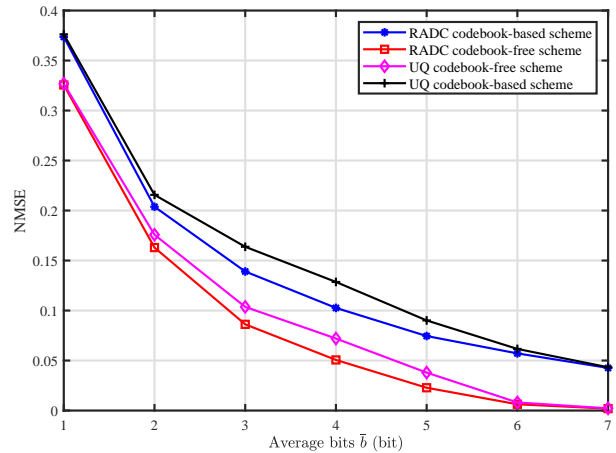}
		\caption{NMSE performance versus the average number of quantization bits $\bar{b}$ ($M=64$, $K=12$, and $\tau=10$).}
        \label{fig_compare_b_avg}
\end{figure}

Fig. \ref{fig_compare_b_avg} compares the NMSE of the proposed RADC
schemes and the UQ schemes (which fundamentally differ in their bit
allocation strategy) as the average number of quantization bits
$\bar{b}$ increases. It can be seen for both codebook-free and
codebook-based cases, the NMSE of the two schemes coincides for small
and large values of $\bar{b}$. However, when $\bar{b}$ (i.e., $3 \leq
\bar{b} \leq5$) is moderate, our proposed schemes significantly
outperform their corresponding UQ scheme. This fact can be explained
as follows: 1) For small values of $\bar{b}$, there is no additional
freedom for adapting the allocation of quantization bits to various
users' propagation conditions; 2) For intermediate values of
$\bar{b}$, thanks to adaptive quantization bit allocation, the
proposed RADC schemes offer added flexibility to select different
resolutions to improve channel estimation accuracy; 3) As $\bar{b}$
becomes sufficiently large, the quantization errors caused by the ADCs
become less important, and no longer represent the main performance
bottleneck of our proposed system.

\begin{figure}[t]
		\centering
		\includegraphics[width=8cm]{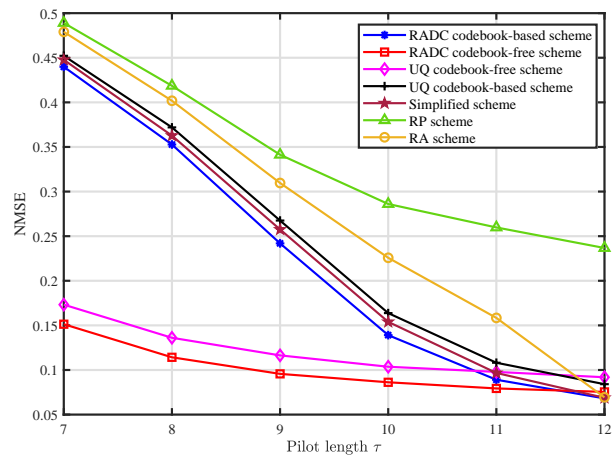}
		\caption{NMSE performance versus the length of pilot sequence $\tau$ ($M=64$, $K=12$, and $\bar{b}=3$).}
        \label{fig_compare_seq_tao}
\end{figure}
The effect of the length of pilot sequences $\tau$ on the NMSE
performance of different schemes is illustrated in
Fig. \ref{fig_compare_seq_tao}. It can be seen that our proposed RADC
codebook-free design yields the best performance among the competing
schemes when $\tau\in[7,11]$, while the RADC codebook-based scheme and the simplified scheme
provide the lowest NMSE (almost identical) when $\tau=12$. This result is expected
because in this case, there are orthogonal pilots to allocate among
all the $K=12$ users, which allows significantly better channel
estimation accuracy. In general, as the pilot length increases, a
noticeable decrease of the channel estimation error can be observed in
both RADC schemes. However, the RADC codebook-free scheme holds
distinct advantages when short pilots are used, which is of crucial
importance for applications with stringent constraints on pilot length.

\begin{figure}[t]
		\centering
		\includegraphics[width=8cm]{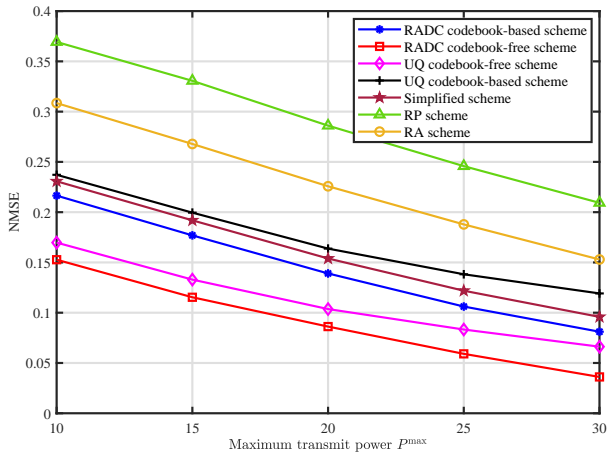}
		\caption{NMSE performance versus the maximum transmit power $P^{\text{max}}$ ($M=64$, $K=12$, $\tau=10$, and $\bar{b}=3$).}
        \label{fig_compare_p}
\end{figure}
Fig. \ref{fig_compare_p} presents the NMSE performance of different schemes versus the maximum transmit power $P^{\text{max}}$. We observe that the NMSE achieved by all schemes is monotonically decreasing with the maximum transmit power. In particular, thanks to the RADC architecture, our proposed RADC codebook-free scheme outperforms its UQ counterpart by a significant margin as the maximum transmit power increases. 
Furthermore, from the figure, the superiority of the proposed RADC codebook-based and simplified schemes are demonstrated once again. These results validate the effectiveness of the proposed design approach based on the channel estimation minimization.

\begin{figure}[t]
		\centering
		\includegraphics[width=8cm]{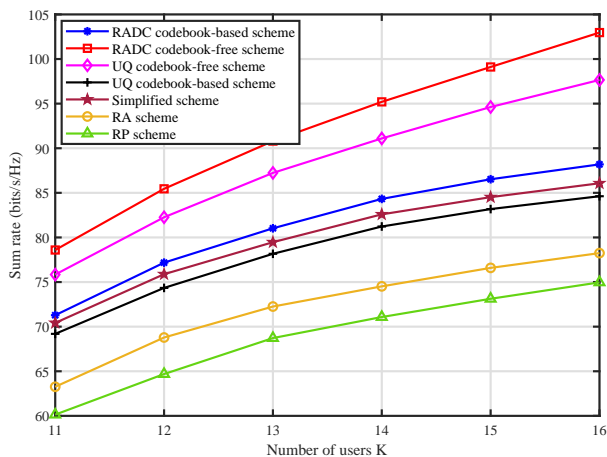}
		\caption{Sum rate performance versus the number users ($M=64$, $\tau=10$, and $\bar{b}=3$).}
        \label{fig_compare_K}
\end{figure}

In Fig. \ref{fig_compare_K}, we investigate the sum rate (bits/s/Hz)
performance of the various schemes, which is evaluated with the aid of
the celebrated precoding algorithm in \cite{wmmse}. As shown in this
figure, our proposed RADC codebook-free scheme outperforms the other
schemes significantly, which confirms its superiority in serving
multiple users. Besides, the RADC codebook-based scheme achieves
better performance in transmission than the simplified scheme and the conventional RA scheme,
showing that it can strike a better trade-off between transmission
rate and complexity.
\begin{figure}[t]
		\centering		\includegraphics[width=8cm]{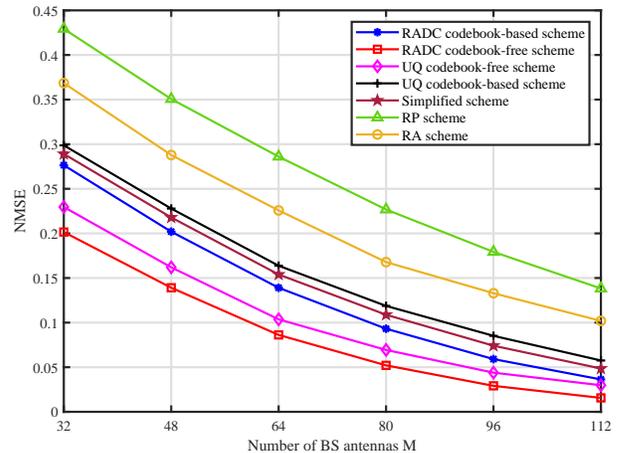}
		\caption{NMSE performance versus the number of antennas at BS ($K=12$, $\tau=10$, and $\bar{b}=3$).}
        \label{fig_compare_M}
\end{figure}

Fig. \ref{fig_compare_M} further shows the NMSE of the channel
estimation versus the number of BS antennas $M$. While the NMSE value
of all the schemes decreases as the number of BS antennas increases,
it is clear that the best performance is achieved by the RADC
codebook-free scheme. Moreover, the performance gap between the
proposed RADC codebook-based and codebook-free schemes shrinks
dramatically as more antennas are being added at the BS.  Hence, the
proposed joint algorithms for pilot sequence design, bit allocation
and hybrid combiner optimization are especially suitable for use in
massive MIMO systems with a large number of antennas $M$. The reason is that our proposed scheme can provide significant flexibility over the ADCs with different channel gains and the joint design framework can exploit the difference in channel quality among links for mitigating the multiuser interference as $M$ increases, thereby supporting more favorable uplink training and channel estimation in a cost-effective manner.
\begin{figure}[t]
		\centering
		\includegraphics[width=8cm]{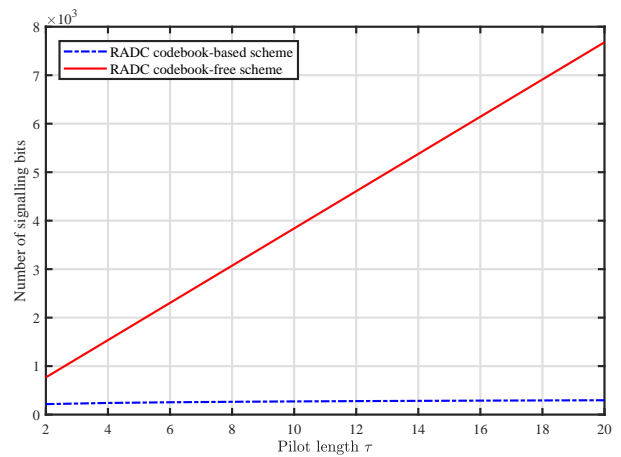}
		\caption{The number of signalling bits versus the length of pilot sequence $\tau$ ($K=24$).}
        \label{fig_compare_signalling}
\end{figure}

Next, we compare the number of required signalling bits for the
feedback of the optimal pilot sequences, for the proposed RADC
codebook-based and codebook-free schemes. Let $B$ denotes the number
of quantization bits for each element of the codebook-free pilot
matrix and of the power vector in the codebook-based pilot scheme. Thus, the
number of signalling bits of the RADC codebook-free scheme is given by
$2BK\tau$ and that of the RADC codebook-based scheme is
$(\lceil\log_{2}\tau\rceil+B)K$, where $\lceil\log_{2}\tau\rceil$
denotes the number of signalling bits required for the codebook
feedback of each user. Fig. \ref{fig_compare_signalling} illustrates
the number of signalling bits versus the length $\tau$ of the pilot
sequences, where we employ $B = 8$ and $K=24$. We can see that the
proposed RADC codebook-based scheme can significantly reduce the
system feedback overhead compared to the RADC codebook-free scheme. In
particular, signalling overhead gap between these two schemes enlarges
with the increase of pilot length.
\begin{table*}[h]
\small
  \caption{Comparison of Average Mutual Coherence of the Optimal Pilot Sequences in Codebook-Free Schemes.}
  \label{compare_benefit}
  \centering
  \begin{tabular}{|c|c|c|c|c|c|c|c|c|c|c|c|c|}
  \hline
  \text{User}&1&2&3&4&5&6&7&8&9&10&11&12\\
  \hline
  \text{RADC Scheme} &0.1327&0.2018&0.2151&0.3563&0.4029&0.5501&0.5322&0.6041&0.6866&0.6736&0.6898&0.6905\\
  \hline
  \text{UQ Scheme}&0.2731&0.5068&0.5626&0.5906&0.6538&0.6282&0.7356&0.7514&0.7673&0.7765&0.7814&0.7818\\
  \hline
  \end{tabular}
\end{table*}

The average mutual coherence of different users' pilot sequences
obtained in the RADC codebook-free and UQ codebook-free schemes is
shown in Table \ref{compare_benefit}. We use the factor
$\mathrm{c}_{k,i}\triangleq\frac{|\mathbf{s}_{k}^{H}\mathbf{s}_{i}|}{|\mathbf{s}_{k}||\mathbf{s}_{i}|}$
as a measure of the mutual coherence, which reveals the degree of
orthogonality among pilot sequences of different users. We label the
users based on the strength of channel pathloss, specifically: user 1
encounters the largest pathloss and user 12 encounters the smallest
one. It is observed from Table \ref{compare_benefit} that both the
RADC codebook-free and UQ codebook-free schemes tend to allocate more
orthogonal pilots to the users with large pathloss, while assigning
less orthogonal pilots to the users with small pathloss. This is
reasonable since users with weak channel gain are more easily affected
by interference. These results show that codebook-free schemes can
take advantage of the knowledge of statistical CSI to in online pilot
design in order to enhance the accuracy of channel estimation.


\section{Conclusions}
In this paper, we investigated the problem of channel estimation in
the uplink massive MIMO systems using RADCs at the BS. We aimed for
minimizing the MSE of the Rayleigh fading channel estimates by jointly
optimizing the pilot sequences, HAD combiners, and the allocation of
ADC quantization bits under practical constraints. To solve such a
challenging nonconvex problem, we harnessed the FP technique and
introduced some auxiliary variables for transforming the original
problem into an equivalent but more manageable form. Then, we
developed new BCD-based and PDD-based algorithms for solving the
resultant equivalent problem for codebook-free and codebook-based
pilot schemes, respectively. Furthermore, we proposed a simplified algorithm for the codebook-based
pilot scheme with much reduced complexity. Our simulation results demonstrated the
efficiency of the proposed algorithms and their superiority in terms
of the MSE and sum rate over the benchmark schemes. It was shown that
the RADC codebook-free scheme generally provides better performance
than the RADC codebook-based scheme, although the latter entails lower
feedback overhead. Hence, the RADC codebook-based scheme is
particularly suitable for application scenarios associated with low
overhead requirement, while the RADC codebook-free scheme is
recommended for applications requiring support for large-scale user
access, high channel estimation accuracy and relaxed overhead
requirements.
\appendix
\subsection{Computation of $\hat{\mathbf{h}}_{k}$ in (\ref{channel estimation}) and $\mathrm{MSE}_{k}$ in (\ref{mse})}
The estimate $\hat{\mathbf{h}}_{k}$ of the original channel $\mathbf{h}_{k}$ is obtained by means of
the MMSE estimation method based on the observation of $\mathbf{y}_{k}$. Hence, it follows from the standard result in estimation theory \cite{estimate} that:
{\setlength\abovedisplayskip{0mm} \setlength\belowdisplayskip{0mm}
\begin{equation}\label{computation}
\hat{\mathbf{h}}_{k}=\mathbb{E}[\mathbf{h}_{k}\mathrm{vec}(\mathbf{y}_{k})^{H}](\mathbb{E}[\mathrm{vec}(\mathbf{y}_{k})\mathrm{vec}(\mathbf{y}_{k})^{H}])^{-1}\mathrm{vec}(\mathbf{y}_{k}).
\end{equation}}
Substituting  (\ref{vec_h}) into (\ref{computation}), and after some mathematical manipulations, we obtain $\hat{\mathbf{h}}_{k}$ in (\ref{channel estimation}).
Denoting the channel estimation error at the BS as $\boldsymbol\varpi_{k}=\hat{\mathbf{h}}_{k}-\mathbf{h}_{k}$, the corresponding $\mathrm{MSE}_{k}$ is given by $\mathrm{MSE}_{k}=\mathbb{E}[\|\boldsymbol\varpi_{k}\|^{2}]=\mathbb{E}[\|\hat{\mathbf{h}}_{k}-\mathbf{h}_{k}\|^{2}]$.
Thus, using (\ref{vec_h}), (\ref{channel estimation}) and after some mathematical manipulations, one can obtain
{\setlength\abovedisplayskip{0mm} \setlength\belowdisplayskip{0mm}
\begin{equation}
\mathrm{MSE}_{k}=\mathrm{tr}(\mathbf{R}_{k}-\mathbf{A}_{k}^{H}\mathbf{B}_{k}^{-1}\mathbf{A}_{k}),
\end{equation}}
where $\mathbf{A}_{k} = \mathbf{s}_{k}\mathbf{x}_{k}\mathbf{R}_{k}$, and $\mathbf{B}_{k} = \sum_{i}\mathbf{s}_{i}\mathbf{x}_{k}\mathbf{R}_{i}\mathbf{x}_{k}^{H}\mathbf{s}_{i}^{H} + \sigma^{2}\mathbf{x}_{k}\mathbf{x}_{k}^{H}\mathbf{I}_{\tau} + (\mathbf{I}_{\tau}\otimes \mathbf{v}_{k}\mathbf{Q})\mathrm{diag}\Big(\sum_{i}(\mathbf{s}_{i}\otimes \mathbf{U})\mathbf{R}_{i}(\mathbf{s}_{i}\otimes \mathbf{U})^{H}+\sigma^{2}\mathbf{I}_{\tau}\otimes \mathbf{U}\mathbf{U}^{\mathbf{H}}\Big)(\mathbf{I}_{\tau}\otimes \mathbf{v}_{k}^{H})\succ 0$.

\vspace{-0.6em}
\subsection{Proof of the equivalence between problems (\ref{P1}) and (\ref{P2}) and of the solution $\mathbf{\Gamma}_{k}^{\star}$ in (\ref{gamma})}
To prove the equivalence between problem (\ref{P1}) and problem (\ref{P2}), we first introduce\\ $f_{0_{k}}(\mathbf{b},\mathbf{U},\mathbf{V},\mathbf{S},\boldsymbol{\Gamma}_{k})=\!\mathrm{tr}(2\Re\{\mathbf{A}_{k}^{H}\mathbf{\Gamma}_{k}\}\!-\!\mathbf{\Gamma}_{k}^{H}\mathbf{B}_{k}\mathbf{\Gamma}_{k})$. With fixed $\mathbf{b},\mathbf{U},\mathbf{V},\mathbf{S}$ (that is, fixed $\mathbf{A}_{k}, \mathbf{B}_{k}$), $f_{0_{k}}$ is concave over the auxiliary variable $\mathbf{\Gamma}_{k}$ and is also a quadratic function of $\mathbf{\Gamma}_{k}$. By applying the first order optimality condition, we can obtain the closed-form solution of problem (\ref{P2}), written as $\mathbf{\Gamma}_{k}^{\star}\!=\!\mathbf{B}_{k}^{-1}\mathbf{A}_{k}$. Upon substitution of $\mathbf{\Gamma}_{k}^{\star}$ into (\ref{P2}), we arrive at the equivalence
of problem (\ref{P1}) and problem (\ref{P2}).

\setcounter{equation}{53}
\begin{figure*}
\begin{align}
\mu_{nm}&\hat{\mathbf{U}}(n,m)\!=\!\sum_{k}\!\Bigg(\!\!(\bar{\mathbf{\Psi}}(n,n)\!+\!\hat{\mathbf{\Psi}}(n,n))\hat{\mathbf{U}}(n,m)\!+\!\sum_{i}\!(\tilde{\mathbf{\Psi}}(n,n)\!+\!\mathbf{\Psi}_{i}(n,n))\hat{\mathbf{U}}(n,m)\mathbf{R}_{i}(m,m)\!\!\Bigg).\label{mu_u}
\end{align}
\hrulefill
\end{figure*}
\begin{figure*}
\begin{equation}
\varphi_{nm}\!=\!\sum_{k}\!\Bigg(\!(\bar{\mathbf{\Psi}}(n,n)\!+\!\hat{\mathbf{\Psi}}(n,n))\hat{\mathbf{U}}(n,m)\!+\!\sum_{i}(\tilde{\mathbf{\Psi}}(n,n)\!+\!\mathbf{\Psi}_{i}(n,n))\hat{\mathbf{U}}(n,m)\mathbf{R}_{i}(m,m)\!\Bigg)\!+\!\mathbf{\Phi}(n,m)\label{value_varphi}.
\end{equation}
\hrulefill
\end{figure*}
\setcounter{equation}{47}
\subsection{Quadratic optimization problem with constant-modulus constraints for $\mathbf{U}$}
We provide an iterative algorithm to solve the following constant-modulus constrained quadratic optimization problem
{\setlength\abovedisplayskip{0mm} \setlength\belowdisplayskip{0.6mm}
\begin{subequations}\label{sub_U_appendix}
\begin{align}
\underset{\mathbf{U}}{\max}&\quad f_{1}(\tilde{\mathbf{b}},\mathbf{U},\mathbf{V},\mathbf{S},\{\mathbf{\Gamma}_{k}\}).\\
\mbox{s.t.}&\quad\ |\mathbf{U}(n,m)| = \frac{1}{\sqrt{M}},\forall{n,m}.
\end{align}
\end{subequations}}
By introducing $\mathbf{M}=\mathrm{diag}(\mathbf{v}_{k}^{H})$, the cost function $f_{1}(\tilde{\mathbf{b}},\mathbf{U},\mathbf{V},\mathbf{S},\{\mathbf{\Gamma}_{k}\})$ can be rewritten as
\begin{align}
f_{1}&(\tilde{\mathbf{b}},\mathbf{U},\mathbf{V},\mathbf{S},\{\mathbf{\Gamma}_{k}\})\nonumber\\
=&\sum_{k}\bigg(\mathrm{tr}(2\Re\{\mathbf{R}_{k}\mathbf{x}_{k}^{H}\mathbf{s}_{k}^{H}\mathbf{\Gamma}_{k}\})-\sum_{i}\mathrm{tr}(\mathbf{\Gamma}_{k}^{H}\mathbf{s}_{i}\mathbf{x}_{k}\mathbf{R}_{i}\mathbf{x}_{k}^{H}\mathbf{s}_{i}^{H}\mathbf{\Gamma}_{k})\nonumber\\ &-\sum_{i}\mathrm{tr}(\mathrm{diag}(\mathbf{s}_{k}\mathbf{s}_{k}^{H})\mathbf{\Gamma}_{k}\mathbf{\Gamma}_{k}^{H})\mathrm{tr}(\mathbf{Q}\mathbf{U}\mathbf{R}_{i}\mathbf{U}^{H}\mathbf{M}\mathbf{M}^{H})\nonumber\\
&-\sigma^{2}\mathrm{tr}(\mathbf{\Gamma}_{k}\mathbf{\Gamma}_{k}^{H})(\mathrm{tr}(\mathbf{Q}\mathbf{U}\mathbf{U}^{H}\mathbf{M}\mathbf{M}^{H})+\mathbf{x}_{k}\mathbf{x}_{k}^{H})\bigg).
\end{align}
We utilize the BCD-type algorithm to tackle problem (\ref{sub_U_appendix}), which is guaranteed to converge to a stationary solution \cite{guarantee}. Specifically, we update each entry of $\mathbf{U}$ once at a time, while keeping the other entries fixed in each step.
The function $f_{1}(\tilde{\mathbf{b}},\mathbf{U},\mathbf{V},\mathbf{S},\{\mathbf{\Gamma}_{k}\})$ restricted to a particular entry $\mathbf{U}(n,m)$ can be represented as a quadratic function of $\mathbf{U}(n,m)$ in the form of
$\hat{f}(\mathbf{U}(n,m))= 2\Re\{\varphi_{nm}^{*}\mathbf{U}(n,m)\}-\mu_{nm}|\mathbf{U}(n,m)|^{2}$,
for some complex number $\varphi_{nm}$ and real number $\mu_{nm}$.
Then, the problem of maximizing $\hat{f}(\mathbf{U}(n,m))$ with respect to $\mathbf{U}(n,m)$ subject to the constant-modulus constraint is given by
\begin{equation}\label{U_nm}
\underset{|\mathbf{U}(n,m)| = \frac{1}{\sqrt{M}}}{\max}\quad\hat{f}(\mathbf{U}(n,m)). \\
\end{equation}
Considering that $|\mathbf{U}(n,m)|=1/\sqrt{M}$, problem (\ref{U_nm}) reduces to
\begin{equation}\label{U_nm1}
\underset{|\mathbf{U}(n,m)| = \frac{1}{\sqrt{M}}}{\max} \Re\{\varphi_{nm}^{*}\mathbf{U}(n,m)\}. \\
\end{equation}
It follows that the optimal solution of $\mathbf{U}(n,m)$ is given by $\varphi_{nm}/\sqrt{M}|\varphi_{nm}|$. Apparently, we only need to know the value of $\varphi_{nm}$ when updating $\mathbf{U}(n,m)$; below, we show how to obtain $\varphi_{nm}$. On the one hand, we have \cite{matrix}
\begin{equation}\label{partial1}
\left.\frac{\partial{\hat{f}(\mathbf{U}(n,m))}}{\partial{\mathbf{U}^{*}(n,m)}}\right|_{\mathbf{U}(n,m)=\hat{\mathbf{U}}(n,m)}=\varphi_{nm}-\mu_{nm}\hat{\mathbf{U}}(n,m).
\end{equation}
On the other hand, let us introduce the following matrix \cite{matrix}
\begin{align}\label{partial2}
\mathbf{\Phi}&=\left.\frac{\partial{f_{1}(\tilde{\mathbf{b}},\mathbf{U},\mathbf{V},\mathbf{S},\{\mathbf{\Gamma}_{k}\})}}{\partial{\mathbf{U}^{*}}}\right|_{\mathbf{U}\!=\!\hat{\mathbf{U}}}\nonumber\\&\!=\!\!\sum_{k}\!\Big(\!\mathbf{Q}_{\alpha}\mathbf{v}_{k}^{H}\mathbf{s}_{k}^{H}\mathbf{\Gamma}_{k}\mathbf{R}_{k}\!\Big)\!\!-\!\!\sum_{k}\!\Big((\bar{\mathbf{\Psi}}\!+\!\hat{\mathbf{\Psi}})\hat{\mathbf{U}}\!+\!\sum_{i}\!(\tilde{\mathbf{\Psi}}\!+\!\mathbf{\Psi}_{i})\hat{\mathbf{U}}\mathbf{R}_{i}\Big),
\end{align}
where $\bar{\mathbf{\Psi}}\!\triangleq\!\sigma^{2}\mathrm{tr}(\mathbf{\Gamma}_{k}\mathbf{\Gamma}_{k}^{H})\mathbf{Q}_{\alpha}\mathbf{v}_{k}^{H}\mathbf{v}_{k}\mathbf{Q}_{\alpha}$, $\hat{\mathbf{\Psi}}\!\triangleq\!\sigma^{2}\mathrm{tr}(\mathbf{\Gamma}_{k}\mathbf{\Gamma}_{k}^{H})\mathbf{M}\mathbf{M}^{H}\mathbf{Q}$, $\tilde{\mathbf{\Psi}}\!\triangleq\!\mathrm{tr}(\mathrm{diag}(\mathbf{s}_{k}\mathbf{s}_{k}^{H})\mathbf{\Gamma}_{k}\mathbf{\Gamma}_{k}^{H})\mathbf{M}\mathbf{M}^{H}\mathbf{Q}$, and $\mathbf{\Psi}_{i}\triangleq\mathbf{Q}_{\alpha}\mathbf{v}_{k}^{H}\mathbf{s}_{i}^{H}\mathbf{\Gamma}_{k}\mathbf{\Gamma}_{k}^{H}\mathbf{s}_{i}\mathbf{v}_{k}\mathbf{Q}_{\alpha}$.
Combining (\ref{partial1}) and (\ref{partial2}), we obtain $\mathbf{\Phi}(n,m)=\varphi_{nm}-\mu_{nm}\hat{\mathbf{U}}(n,m)$. By expanding $[(\bar{\mathbf{\Psi}}\!+\!\hat{\mathbf{\Psi}})\hat{\mathbf{U}}\!+\!\!\sum_{i}(\tilde{\mathbf{\Psi}}+\mathbf{\Psi}_{i})\hat{\mathbf{U}}\mathbf{R}_{i}](n,m)$ and examining the coefficient of $\hat{\mathbf{U}}(n,m)$, we find (\ref{mu_u}) displayed at the top of this page.
Therefore, the value of $\varphi_{nm}$ is determined as (\ref{value_varphi}), shown at the top of this page.

\vspace{-1.1em}
\subsection{Proof of Theorem \ref{theorem1}}
We focus on the proof of Theorem \ref{theorem1}, i.e., the convergence of Algorithm \ref{algorithm_non}. It is seen that each subproblem in the proposed BCD-based Algorithm 1 is guaranteed to converge to its stationary point. Specifically, the subproblem with respect to $\mathbf{U}$, $\mathbf{V}$ and $\mathbf{S}$ can be globally solved, respectively, and the solutions satisfy the optimality conditions. As for the subproblem for $\tilde{\mathbf{b}}$ (the relaxation variable of $\mathbf{b}$), we leverage the SCA method and construct the concave surrogate function $\bar{f}^{t}(\tilde{\mathbf{b}})$ to approximate the nonconcave objective function $f_{1}(\tilde{\mathbf{b}})$. Then, the complex nonconcave optimization subproblem (\ref{sub_b}) is transformed into a concave subproblem (\ref{sub_b1}) with guaranteed convergence. The valid surrogate function $\bar{f}^{t}(\tilde{\mathbf{b}})$ and $f_{1}(\tilde{\mathbf{b}})$ have the same value and gradient at point $\tilde{\mathbf{b}}^{t}$ which follows from the SCA theory \cite{sca}. Consequently, subproblem (\ref{sub_b}) and subproblem (\ref{sub_b1}) will share the same stationary point.

With objective function value of each subproblem nondecreasing and guaranteed to converge to its stationary point and according to Proposition 2.7.1 (Convergence of BCD) in \cite{guarantee}, Algorithm 1 is guaranteed to converge to a stationary point of problem (\ref{P2}). Furthermore, based on Lemma 1, problem (\ref{P1}) and problem (\ref{P2}) can share the same stationary points. Therefore, Algorithm 1 is guaranteed to converge to a stationary point of problem (\ref{P1}).


\end{document}